%% file: ms.tex
\documentclass[12pt,preprint]{aastex}
\usepackage{apjfonts}
\usepackage{color}

\def\wig#1{\mathrel{\hbox{\hbox to 0pt{%
          \lower.5ex\hbox{$\sim$}\hss}\raise.4ex\hbox{$#1$}}}}

\shorttitle{Thermal Evolution of the Giant Planets}
\shortauthors{Fortney, et al.}

\newcommand{\rj}{$R_{\mathrm{J}}$}
\newcommand{\me}{$M_{\oplus}$}
\newcommand{\re}{$R_{\oplus}$}

\newcommand{\ls}{$L_{\odot}$}

\newcommand{\te}{$T_{\rm eff}$}
\newcommand{\teff}{$T_{\rm eff}$}
\newcommand{\teq}{$T_{\rm eq}$}

\newcommand{\ti}{$T_{\rm int}$}
\newcommand{\tten}{$T_{10}$}
\newcommand{\tone}{$T_{1}$}
\newcommand{\cp}{\citep}
\newcommand{\ct}{\citet}

\slugcomment{Astrophysical Journal, in press}

\begin{document}

\title{Self-Consistent Model Atmospheres and the Cooling of the Solar System's\\Giant Planets}

\author{J. J. Fortney\altaffilmark{1}$^{,}$\altaffilmark{2}, M. Ikoma\altaffilmark{3}, N. Nettelmann\altaffilmark{1}, T. Guillot\altaffilmark{4}, M. S. Marley\altaffilmark{5}} 

\altaffiltext{1}{Department of Astronomy and Astrophysics, University of California, Santa Cruz, CA 95064; jfortney@ucolick.org}
\altaffiltext{2}{Alfred P. Sloan Research Fellow}
\altaffiltext{3}{Department of Earth and Planetary Sciences, Tokyo Institute of Technology, Ookayama, Meguro-ku, Tokyo 152-8551, Japan}
\altaffiltext{4}{Observatoire de la C\^ote d'Azur, Laboratoire Cassiop\'ee, CNRS UMR 6202, BP 4229, 06304 Nice cedex 4, France}
\altaffiltext{5}{Space Science and Astrobiology Division, NASA Ames Research Center, Mail Stop 245-3, Moffett Field, CA 94035}

\begin{abstract}
We compute grids of radiative-convective model atmospheres for Jupiter, Saturn, Uranus, and Neptune over a range of intrinsic fluxes and surface gravities.  The atmosphere grids serve as an upper boundary condition for models of the thermal evolution of the planets.  Unlike previous work, we customize these grids for the specific properties of each planet, including the appropriate chemical abundances and incident fluxes as a function of solar system age.  Using these grids, we compute new models of the thermal evolution of the major planets in an attempt to match their measured luminosities at their known ages.  Compared to previous work, we find longer cooling times, predominantly due to higher atmospheric opacity at young ages.  For all planets, we employ simple ``standard'' cooling models that feature adiabatic temperature gradients in the interior H/He and water layers, and an initially hot starting point for the calculation of subsequent cooling.  For Jupiter we find a model cooling age $\sim$10\% longer than previous work, a modest quantitative difference.  This may indicate that the hydrogen equation of state used here overestimates the temperatures in the deep interior of the planet.  For Saturn we find a model cooling age $\sim$20\% longer than previous work.  However, an additional energy source, such as that due to helium phase separation, is still clearly needed.  For Neptune, unlike in work from the 1980s and 1990s, we match the measured \teff\ of the planet with a model that also matches the planet's current gravity field constraints.  This is predominantly due to advances in the high-pressure equation of state of water.  This may indicate that the planet possesses no barriers to efficient convection in its deep interior.  However, for Uranus, our models exacerbate the well-known problem that Uranus is far cooler than calculations predict, which could imply strong barriers to interior convective cooling.  The atmosphere grids are published here as tables, so that they may be used by the wider community.

\end{abstract}


\section{Introduction}
Planets cool as they age.  For giant planets, what regulates the cooling of their mostly convective interiors is the radiative properties of the thin skin of atmosphere that rests atop the bulk of the planet's mass.  The effect of atmospheres on giant planet cooling was first investigated in the mid 1970s, when \ct{Graboske75}, \ct{Hubbard77}, and \ct{Pollack77} computed the first thermal evolution models of Jupiter and Saturn that coupled the atmospheric and interior structures.

The relationship between the atmosphere and interior has become even better appreciated since the discovery of hot Jupiters 15 years ago \cp{Mayor95}, which orbit their parent stars at 5-10 stellar radii, and often intercept $10^4-10^5$ more flux than Jupiter receives from the Sun.  This incident flux drives the external radiative zone to pressures near 1 kbar in Gyr-old planets \cp{Guillot96,Sudar03,Fortney07a}, which suppresses the transfer of intrinsic flux through the atmosphere, thereby slowing interior cooling and contraction.  A great variety of model atmosphere grids have been calculated for these close-in planets \cp[e.g.][]{Burrows03,Baraffe03,Fortney07a,Baraffe08,Ibgui09}, which feature state-of-the-art treatments of chemistry \cp[e.g.][]{Lodders06}, non-gray opacities \cp[e.g.][]{Sharp07,Freedman08}, and radiative transfer.

These same atmosphere models have been honed over the past 10 years on excellent spectra of hundreds of brown dwarfs with \te\ down to $\sim$550 K \cp{Allard01,Marley02,Burrows06,Saumon06,Saumon08}.  But these tools have not been applied to generate modern model atmosphere grids for use in evolution models of the solar system's giant planets.

The first, and last, grid of non-gray radiative-convective model atmospheres to be tabulated in print for use in evolutionary calculations were presented in \ct{Graboske75}.  These were computed for planetary effective temperatures of 20 to 1900 K, at only two surface gravities, that of the current Jupiter, and at 1/64 Jupiter's current gravity.  More recently, \ct{Burrows97} and \ct{Baraffe03} computed non-gray atmosphere models for use in the evolution of brown dwarfs and giant planets, including Jupiter and Saturn \cp{Hubbard99,FH03}, but these grids are not available to the wider community.  Both \ct{Graboske75} and \ct{Burrows97} did not treat irradiation from the Sun.

The purpose of the present paper is to provide radiative-convective model atmosphere grids for Jupiter, Saturn, Uranus, and Neptune, for the full ranges of surface gravities, effective temperatures, incident fluxes, and atmospheric compositions specific to these planets, as a function of planet age.  This will greatly minimize any remaining uncertainty in cooling calculations related to the atmospheric boundary condition for these planets.  Here we also apply these grids to investigate what effect they have on the cooling histories of the planets.

We can briefly summarize the main findings of evolution models of the major planets\footnote{A nice review of work until 1980, the early years of giant planet spacecraft observations and evolution modeling, can be found in \ct{Hubbard80a}}.  Going back to \ct{Hubbard77}, evolution models of Jupiter have yielded a planetary \te\ roughly consistent with observations at an age of 4.6 Gyr, under the assumption of a fully adiabatic interior \cp{Hubbard99, FH03}, or with a small radiative zone \cp{Guillot95}.  These models use an initial, post-formation \teff\ that is arbitrarily large \cp[see e.g.][]{Marley07}, on the order of 1000 K, implying that Jupiter was quite luminous at young ages, as is expected from any model of its formation.  Jupiter's evolution is considered fairly well understood.

Fully convective homogeneous models of Saturn reach the planet's known \te\ in only 2-3 Gyr, implying that Saturn is much warmer than can be accommodated by the same kind of model that works well for Jupiter \cp{Pollack77}.  The phase separation of helium from liquid metallic hydrogen, and its subsequent sinking to deeper layers (a differentiation process) has long been suspected of providing Saturn's additional energy source \cp{Salpeter73,Stevenson75,SS77a,SS77b}.  While this is widely assumed to be true, the exact details of the physics \cp{Morales09,Lorenzen09} and its implementation in cooling models \cp{Hubbard99,FH03} has not yet been solved.

Uranus and Neptune cooling models have their own problems, which are different than those of Jupiter and Saturn.  \ct{Hubbard78} and \ct{Hubbard80b} computed the first thermal evolution models of these planets.  \ct{Hubbard80b} could only find good agreement between model cooling ages and the age of the solar system, if the planets started out relatively cool, with post-formation \teff s only $\sim$~50\% greater than their current values.  Put another way, the current planets were found to be underluminous compared to models that started with the arbitrarily large initial \teff\ values used for Jupiter and Saturn.  The problem was especially pronounced for Uranus.

Later work on cooling models can be found in \ct{Ubook} and \ct{Nbook}.  Again, both Uranus and Neptune appeared \emph{under-luminous} compared to models, although the discrepancy was larger for Uranus than for Neptune.  \ct{Ubook} hypothesized that statically stable layers in the interior of the ``ice giants,'' due to compositional layering, could lock in internal energy, leading to small intrinsic fluxes.  They suggested that only the outer 40\% and 60\% in radius of Uranus and Neptune, respectively, were freely convecting.  It is known that the interior structures of these planets do not need to be partitioned into well-defined layers \cp{Marley95}.  More recently, and importantly, \ct{Fortney09} computed new cooling models of these planets.  They used a new water equation of state \cp{French09}, and find that Neptune's current \te\ can be matched, while the problem with Uranus remains, for models that feature the standard arbitrary hot start.  The implication is that the argument for stable layers in Neptune may no longer be necessary.

To summarize, current state-of-the-art models can reproduce the \te\ of Jupiter and Neptune, while Saturn is over-luminous and Uranus is under-luminous.  In this paper, in \S2 we describe new model atmospheres for all of these planets, and apply them to new cooling models in \S3.  In \S4 we give our conclusions and suggest future work.

\section{Model Atmospheres}
\subsection{The model atmosphere code and validation}
An evolution calculation is dependent on a relation between the intrinsic effective temperature \ti, surface gravity $g$, and specific entropy of the interior adiabat $S$ \cp{Graboske75,Hubbard77,Burrows97}.  Here \ti\ as defined as the \te\ that the planet would have in the absence of solar insolation.  With a third temperature, \teq, the \teff\ that the planet would have in the absence of internal energy, we have the relation \te$^4$=\ti$^4$+\teq$^4$. Sometimes $S$ is parameterized as $T_{10}$, the temperature of the interior adiabat (which is isentropic) at $P=10$ bars \cp{Burrows97,Hubbard99}.  In practice, model atmospheres are calculated at a range of \ti\ and $g$ points and the value of the \teff\ and $T_{10}$ are tabulated for each of these converged models.  This collection of points in then inverted, such that for a given $T_{10}$ and $g$ of a structural model, the \ti\ and \teff\ are found by interpolation.  While \teff\ can be most easily derived from observations, it is \ti\ that is the parameterization of the flux from the planet's interior cooling.  %Mathematically, the cooling calculation can be described by: 
%
%\begin{equation} \label{e1}
%L(t) \equiv 4\pi\sigma R(t)^2 T_{\rm int}(t)^4,
%\end{equation}
%
%and the time-rate of change of $L$ described by
%
%\begin{equation} \label{e2}
%{\partial L \over \partial m} = -T(m) {\partial S(m) \over \partial t},
%\end{equation}
%where
%\begin{equation}
%\label{shell}
%m=\frac{1}{M}\int_{0}^{r} 4 \pi r^{{\prime}^2} dr^{\prime} \rho(r^{\prime}).
%\end{equation}
%Equation \ref{e2} can be rewritten as
%\begin{equation} \label{time}
%\partial t = - \frac{M}{L} \int_{0}^{1}T \partial S dm,
%\end{equation}
%which yields the time step between successive models.

To create the grids we make use of a one-dimensional plane-parallel model atmosphere code that has been widely used for solar system planets, exoplanets, and brown dwarfs over the past two decades.  The optical and thermal infrared radiative transfer solvers are described in detail in \ct{Toon89}.  Past applications of the model include Titan \cp{Mckay89}, Uranus \cp{MM99}, gas giant exoplanets \cp{Marley98,Fortney05,Fortney07b,Marley07b,Fortney08a}, and brown dwarfs \cp{Marley96,Burrows97,Marley02,Saumon08}.  We use the correlated-k method for opacity tabulation \cp{Goody89} over 196 wavelength bins, from 0.26 to 325 $\mu$m for outgoing thermal radiation, and 160 wavelength bins, from 0.26 to 6 $\mu$m, for incident stellar radiation.  Our extensive opacity database is described in \ct{Freedman08}.  We include the opacity of neutral atomic alkalis, which are prominent in brown dwarfs \cp{BMS}, and also close a radiative region in Jupiter's deep atmosphere \cp{Guillot04}.

We make use of tabulations of chemical mixing ratios from equilibrium chemistry calculations of K.~Lodders and collaborators \cp{Lodders99,Lodders02,Visscher06,Lodders06}.  We use the base protosolar abundances of \ct{Lodders03}, with Jupiter atmosphere models enhanced in heavy elements (``metals'') by a factor of $\sim$~3 (specifically [M/H]=+0.5), Saturn by a factor of 10 ([M/H]=+1.0), and Neptune and Uranus by a factor of $\sim$~30 ([M/H]=+1.5).  A 3$\times$ enhancement in Jupiter is roughly consistent with observations by the \emph{Galileo Entry Probe} \cp{Wong04}.  An enhancement of 10$\times$ in Saturn is consistent with the CH$_4$ abundance deduced by \ct{Flasar05} via \emph{Cassini} CIRS spectroscopy.  An enhancement of 30-60$\times$ in Uranus and Neptune is consistent with the methane abundance in these atmospheres \cp[e.g.][]{Guillot07}.

These are the first atmosphere grids for any of these planets to specifically include irradiation from the Sun.  As mentioned, there have been only two previous non-gray atmosphere tabulations applied to evolutionary calculations, \ct{Graboske75} and \ct{Burrows97}, and only the former is publicly available.  Both of these grids assumed solar abundances, no irradiation from the Sun, and were sparsely sampled in \te\ and $g$.

The utilization of no-irradiation atmosphere models rests on the assumption that one can assume that all flux absorbed by the planet is absorbed into the planet's convective zone and directly adds to the energy budget of the planet \cp{Hubbard77,FH03}.  Given that all the giant planets have stratospheres, due to absorption of incident flux in the radiative atmosphere, this assumption does not hold.  In addition, one must assume a Bond albedo at every age, in order to determine the amount of absorbed incident flux.  Typically, the current Bond albedo is used at all ages.  Here we consistently solve for the deposition of stellar flux into the atmosphere as a function of wavelength and depth, and make no assumption regarding the Bond albedos.  The \te\ at a given \ti\ and $g$ is obtained based on the converged radiative-convective equilibrium model atmosphere temperature/pressure/opacity profile\footnote{The Bond albedo cannot be precisely determined from a 1D model.  A multi-D model must be used to investigate the scattering of light over all phase angles \cp[e.g.][]{MM99,Cahoy10}}.  As mentioned above, this improved method of treating the atmospheric boundary has been used for the close-in ``hot Jupiter'' planets for several years \cp[see, e.g.][for initial calculations]{Baraffe03,Burrows03} and was extended by \ct{Fortney07a} to models of EGP thermal evolution for planets from 0.02 to 10 AU.

Here the derived \emph{P--T} profiles are for planet-wide average conditions, with a Sunlight zenith angle of 60 degrees ($\mu$=0.5), and the incident flux cut by 1/2, due to the day/night difference \cp{MM99}.  Using an analytical model for a semi-grey, non-scattering atmosphere \cp{Guillot10}, we compared solutions obtained with that approximation to globally averaged solutions, with a greenhouse factor (i.e. the ratio of the infrared to the visible mean opacities) between $10^{-2}$ to $100$. We found that mean deviations in the temperature profile between optical depths $\tau=10^{-2}$ to $100$ were smaller than $1.4\%$, with a standard deviation smaller than $3.2\%$, i.e. much smaller than other uncertainties. 

While the model atmospheres presented here are a significant advance over previous calculations, some simplifications are still made.  Most importantly, we neglect the opacity of condensate clouds and non-equilibrium hazes.  For instance, the geometric albedos of Jupiter and Saturn are depressed in the blue due to dark hazes \cp{Kark94} that are thought to be derived from the photochemical destruction of methane \cp{Rages99}.  However, to our knowledge there is no adequate theory to predict how the optical depth and distribution of this haze may change as a function of solar luminosity, planetary \teff, and chemical mixing ratios, which change as the planets cool.  Even more importantly, equilibrium condensate clouds such as water or ammonia lead to large geometric albedos outside of methane absorption bands \cp[e.g.][]{Cahoy10}.  However, the cloud optical properties, particle sizes, vertical distribution, and coverage fraction are difficult or impossible to predict from first principles. See \ct{AM01}, \ct{Cooper03}, and \ct{Helling06} for modern cloud models for substellar objects.

Of significant practical importance, and what validates our cloud-free assumption here, is that we reproduce the current temperature structure and \teff\ of the Jovian atmosphere much better with \emph{cloud-free} models than with models that include ammonia and water clouds \cp{Cahoy10}.  The cloudy models (which use the Ackerman \& Marley cloud model) yield model atmospheres that are too reflective (a high Bond albedo) and are hence, too cool.  This is because without dark photochemical hazes, the ammonia and water condensate clouds in the atmosphere model are more reflective than observed.  Cloud layers in Uranus and Neptune reside deeper in the atmosphere, which makes this issue less important for their atmospheres.  Looking back in time, when the planets will be warmer, points to a proper treatment of the time-varying Bond albedos being a very complicated problem. The cloud layers in Uranus and Neptune will be higher in the atmosphere.  Jupiter and Saturn will lose their ammonia clouds, and possess a top layer of thick water clouds.  There is much future work to be done in this area.

\begin{figure}
\plotone{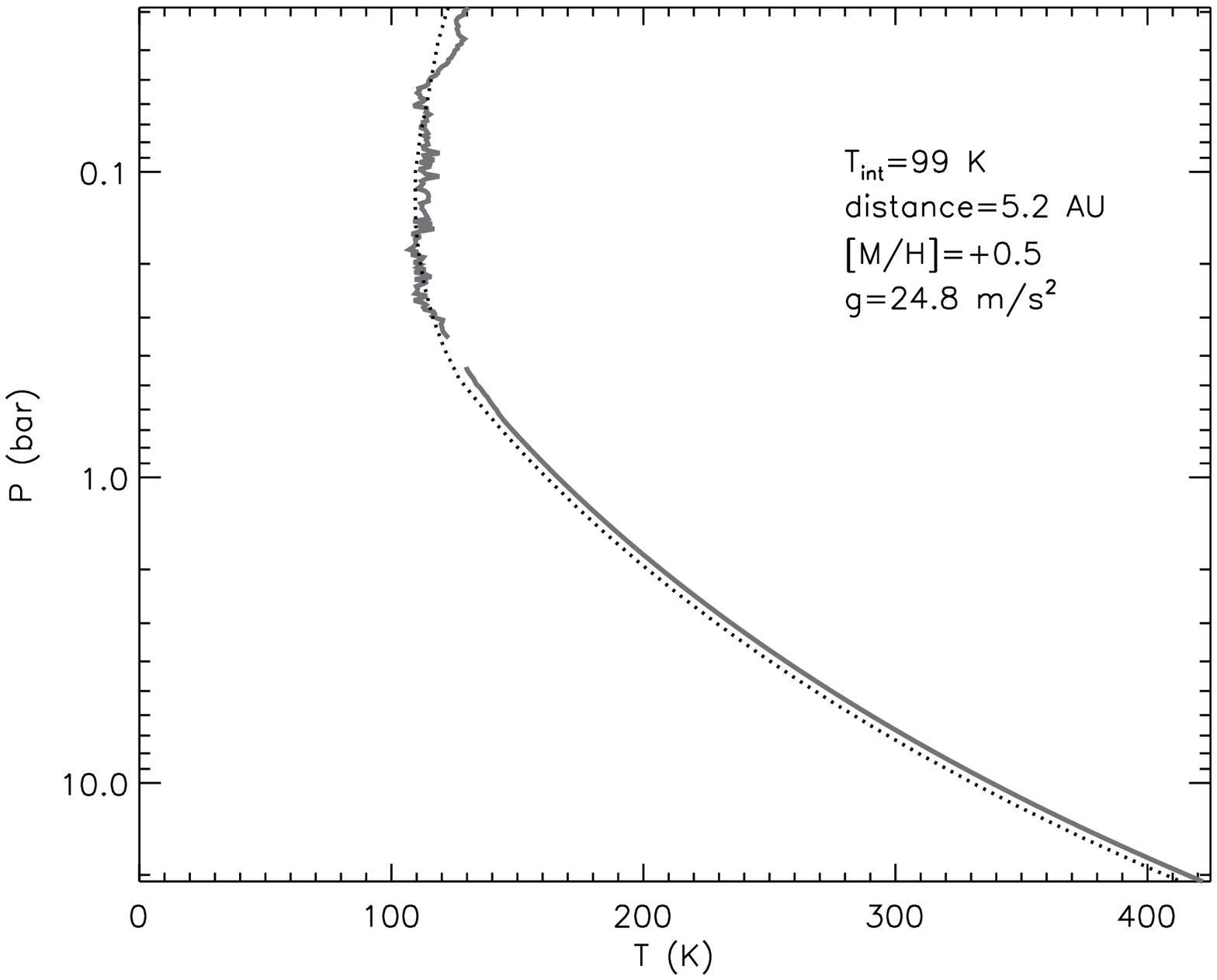}
\caption{Comparison of a model of Jupiter's current equatorial \emph{P-T} profile (dotted black line) to that obtained for Jupiter by the \emph{Galileo Entry Probe} (thick gray lines).  The model parameters are shown on the plot.  The value of \ti\ is chosen to match the value of the intrinsic flux measured by \emph{Voyager 2}.  The resulting \te\ is 124.6 K, the observed value within the observational error bars.  The model planet's 1-bar temperature is cooler by 4 K than that measured by the probe.
\label{gal}}
\end{figure}

\begin{figure}
\plotone{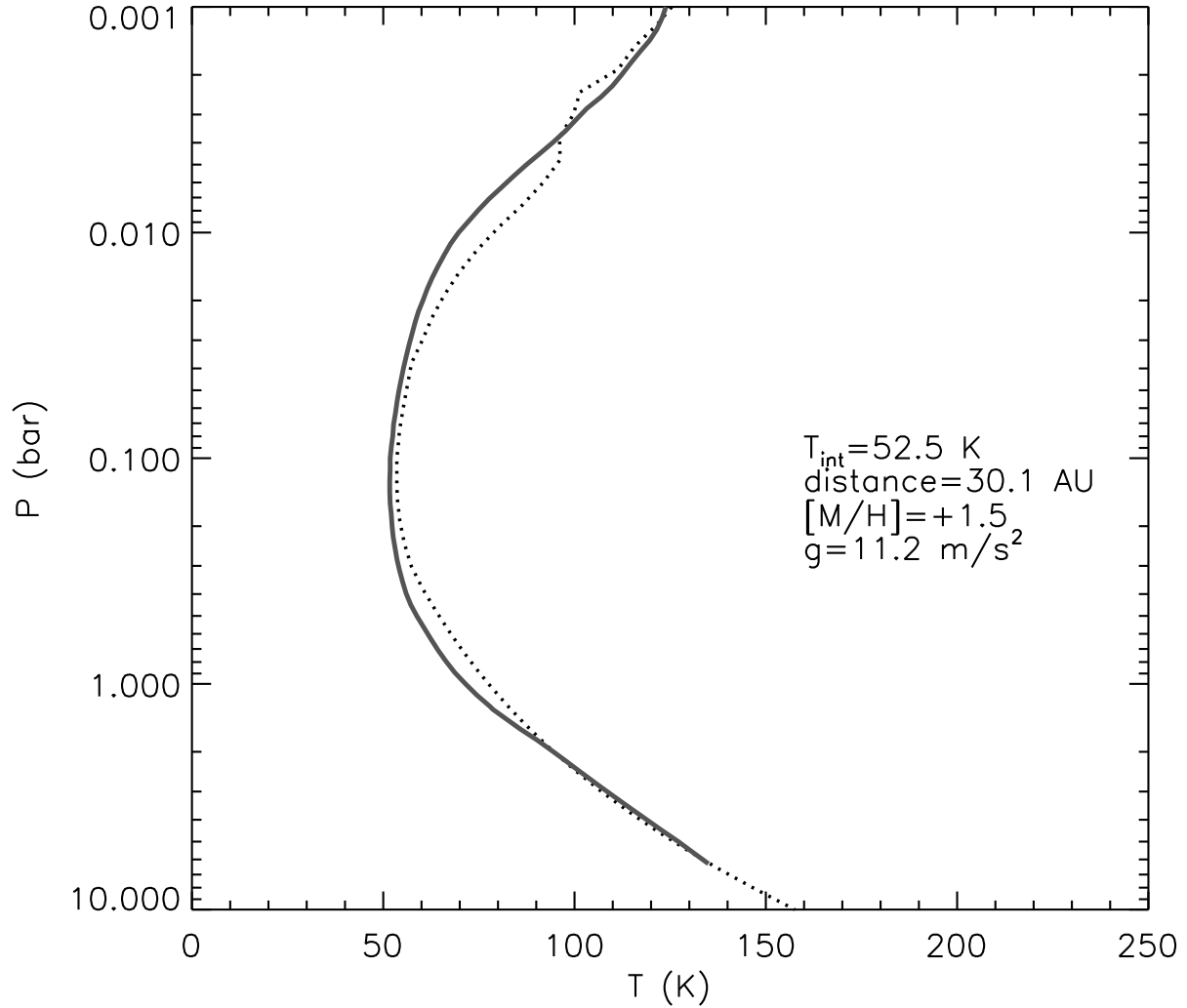}
\caption{Comparison of a model of Neptune's current \emph{P-T} profile (dotted black line) to that obtained for Neptune via radio occultation by \emph{Voyager 2} (thick gray line).  The model parameters are shown on the plot.  The value of \ti\ is chosen to match the value of the intrinsic flux measured by \emph{Voyager 2}.  The resulting \te\ is 59.2 K, the observed value within the observational error bars.  The model planet's 1-bar temperature is warmer by 5 K than that measured by \emph{Voyager 2}.  Extrapolating to the 10-bar level, the temperature appears to be a good match.
\label{neptp}}
\end{figure}

Fortunately, for Jupiter and Saturn, the neglect of both the dark absorbing hazes, and relatively bright condensate clouds somewhat cancel out, and we are able to reproduce the temperature structure of the planets' tropospheres quite well.  The example of a cloud free model of the current Jupiter, compared to data from the \emph{Galileo Entry Probe} \cp{Seiff98}, is shown in Figure \ref{gal}.  The model \te\ is correct to 0.2 K, and the 1-bar temperature (within the convective region) to 4 K.  In Figure \ref{neptp} we show a comparison with the atmosphere of Neptune, which was probed via radio occultation by \emph{Voyager 2} \cp{Lindal92}.  The comparison here is also favorable, with differences only on the order of several degrees.  Since these planets cool very slowly at gigayear ages, a good match to \teff\ and \tten\ now also indicates a good match for the past several gigayears.

\subsection{The atmosphere grids}
\ct{Hubbard77} provided an analytic fit to the original \ct{Graboske75} non-gray (meaning frequency-dependent opacities are used) grid of model atmosphere calculations.  As discussed in \ct{Hubbard80b}, \ct{Guillot95}, and \ct{Saumon96}, this grid can be described by a function of the form
\begin{equation}
T_0=K g^{-0.167} T_{\rm eff}^{1.243}
\end{equation}
where $T_0$ is the temperature at a reference pressure $P_0$ within the convective region of the atmosphere, such as 1 or 10 bars, and $K$ is a constant.  \ct{Saumon96} find that $T_0$=10 bar and K=3.36 provides a reasonable fit below \teff\ of 200 K.  For Uranus and Neptune evolution models, \ct{Hubbard80b}, in order to better match the planet's atmospheres, suppressed the gravity dependence entirely and set $K$=1, with $P_0=0.750$ bar for Uranus, and $P_0=0.861$.  So, in general, a variety of fits, some based on the original calculations, some based on ad-hoc modifications, are available.  Clearly new work in this area is warranted.

The \ct{Burrows97} grid, slightly updated in \ct{Hubbard99}, spans a very wide-range of \te, $g$, and \tten, with a mix of gray atmospheres at high \te, and non-gray atmospheres at low \te, suitable for giant planets.  The \ct{Baraffe03} grid, fully non-gray, is similarly expansive, to treat the lowest mass stars, brown dwarfs, and planets.

\begin{figure}
\plotone{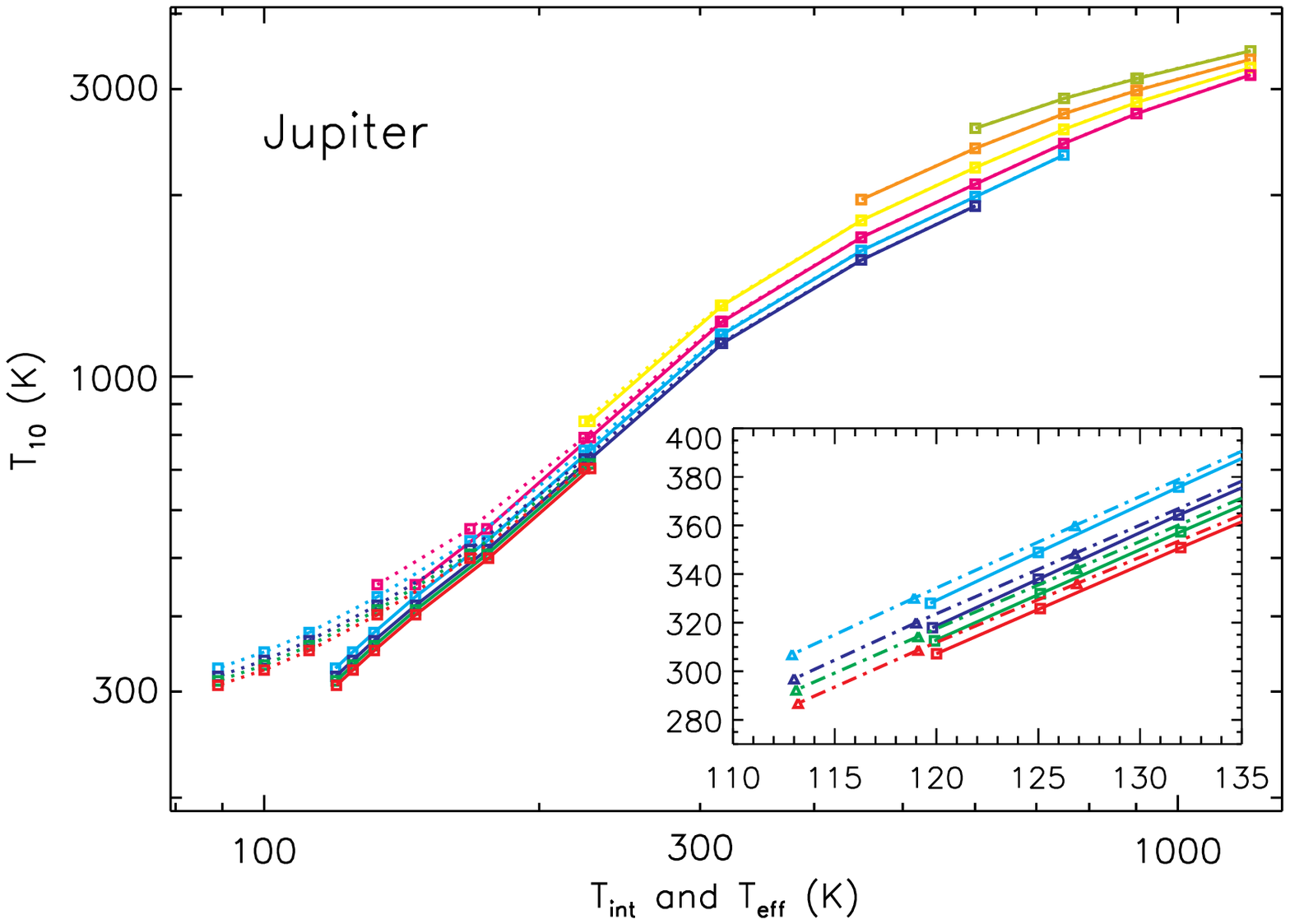}
\caption{The Jupiter model atmosphere grid at 1.0 \ls.  Plotted on the x-axis is the \te\ (solid lines) and \ti\ (dotted lines) vs.~\tten.  Lower gravity models (yellowish-green, orange) cover only high \tten\ and \te.  Higher gravity models (red, bright green) cover only low \tten\ and \te.  The effects of irradiation from the Sun (seen as a higher \te\ than \ti) are only apparent at low temperatures. \emph{Inset:}  \te\ vs.~\tten\ for the 1.0 \ls\ (solid,with squares) and 0.7 \ls\ (dash-dot, with triangles) grids.  This shows that the effects of the lower solar luminosity do have small, but quantitatively important effects at low temperature.
\label{jupgrid}}
\end{figure}

\begin{figure}
\plotone{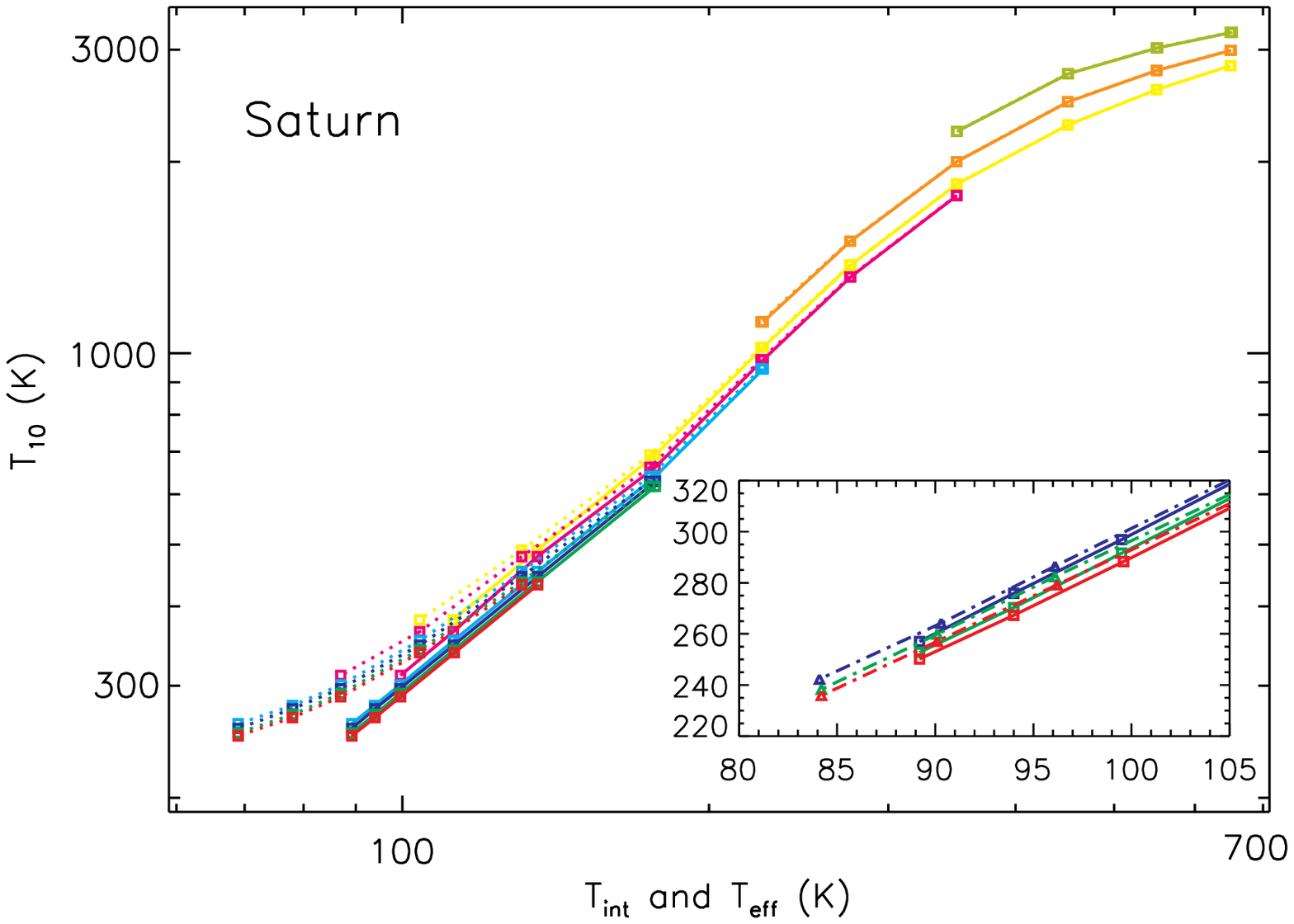}
\caption{The Saturn model atmosphere grid at 1.0 \ls.  Plotted on the x-axis is the \te\ (solid lines) and \ti\ (dotted lines) vs.~\tten.  Lower gravity models (yellowish-green, orange) cover only high \tten\ and \te.  Higher gravity models (bright green, red) cover only low \tten\ and \te.  The effects of irradiation from the Sun (seen as a higher \te\ than \ti) are only apparent at low temperatures.  \emph{Inset:}  \te\ vs.~\tten\ for the 1.0 \ls\ (solid, with squares) and 0.7 \ls\ (dash-dot, with triangles) grids.  This shows that the effects of the lower solar luminosity do have small, but quantitatively important effects at low temperature.
\label{satgrid}}
\end{figure}

We have computed pressure-temperature (\emph{P-T}) profiles for Jupiter and Saturn at $\sim$50 \teff /$g$ pairs, enough for detailed coverage of the evolution of \ti, $g$, and $S$, over their evolution.  For Jupiter, \ti\ ranges from 89 K to 1200 K, while the gravity range covers 0.12 to 1.1 $\times$ Jupiter's gravity.  For Saturn, \ti\ ranges from 69 K to 650 K, while the gravity range covers 0.12 to 1.1 $\times$ Saturn's gravity.  For both planets, these calculations were done at the current value of the incident solar flux, as well as 0.7 of this value so that the evolution of the solar luminosity can be incorporated \cp[e.g.][]{Hubbard99}.  These are included as Tables 1 and 2, for Jupiter and Saturn, respectively. The grids for 1.0 \ls\ for Jupiter and Saturn are shown in Figures \ref{jupgrid} and \ref{satgrid}, respectively.  They show \ti\ and \te\ on the x-axis, and \tten\ on the y-axis, for 8 different surface gravities.  Low gravities cover high \ti\ (young ages) and high gravities cover low \ti\ (old ages).  An inset shows \te\ vs.~\tten\ for the current and lower solar luminosity, at very low \te.  When \ti\ is large, then the planet's energy budget is dominated by its own intrinsic luminosity, and \te\ is only negligibly larger than \ti.  However, at lower values of \ti, eventually there is a clear separation between \ti\ and \te, as expected.

\begin{figure}
\plotone{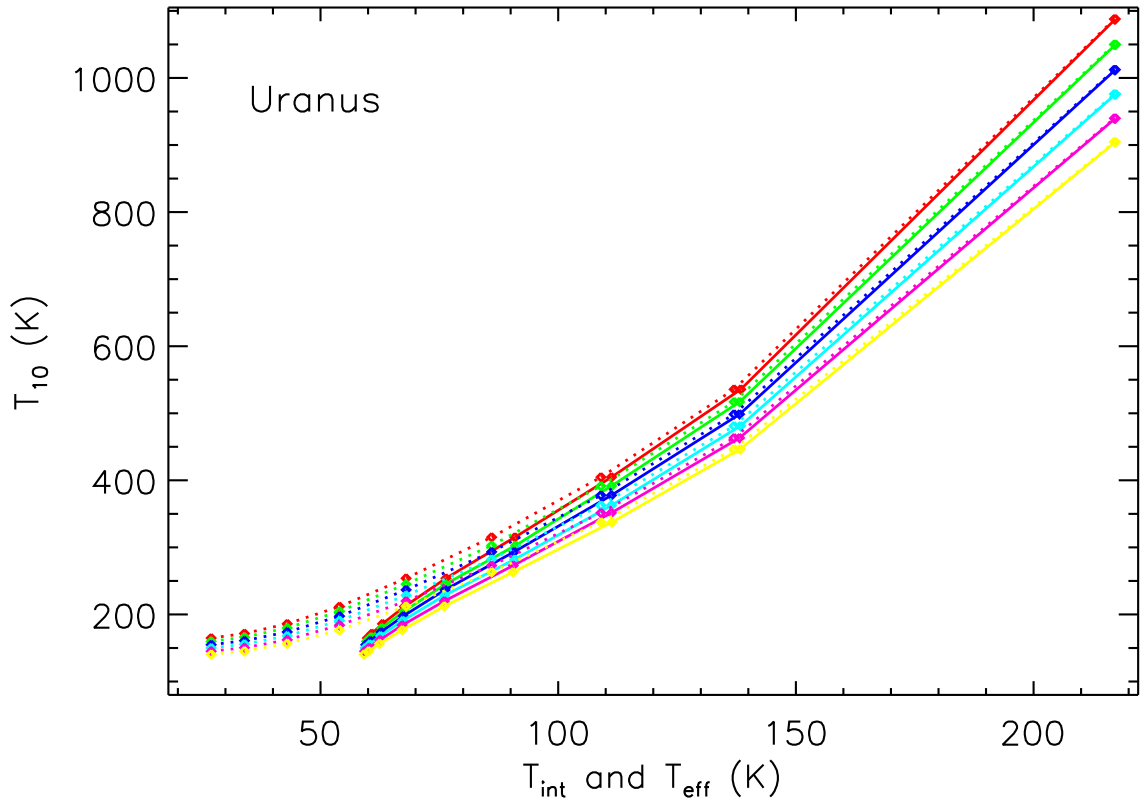}
\caption{The Uranus model atmosphere grid at 1.0 \ls.  Red is the lowest gravity, 4.0 m s$^{-2}$, while yellow is the highest, 12.6 m s$^{-2}$.  The x- and y-axes are the same as in Figures \ref{jupgrid} and \ref{satgrid}, meaning the \te s are solid lines and \ti s are dotted lines.
\label{ugrid}}
\end{figure}

\begin{figure}
\plotone{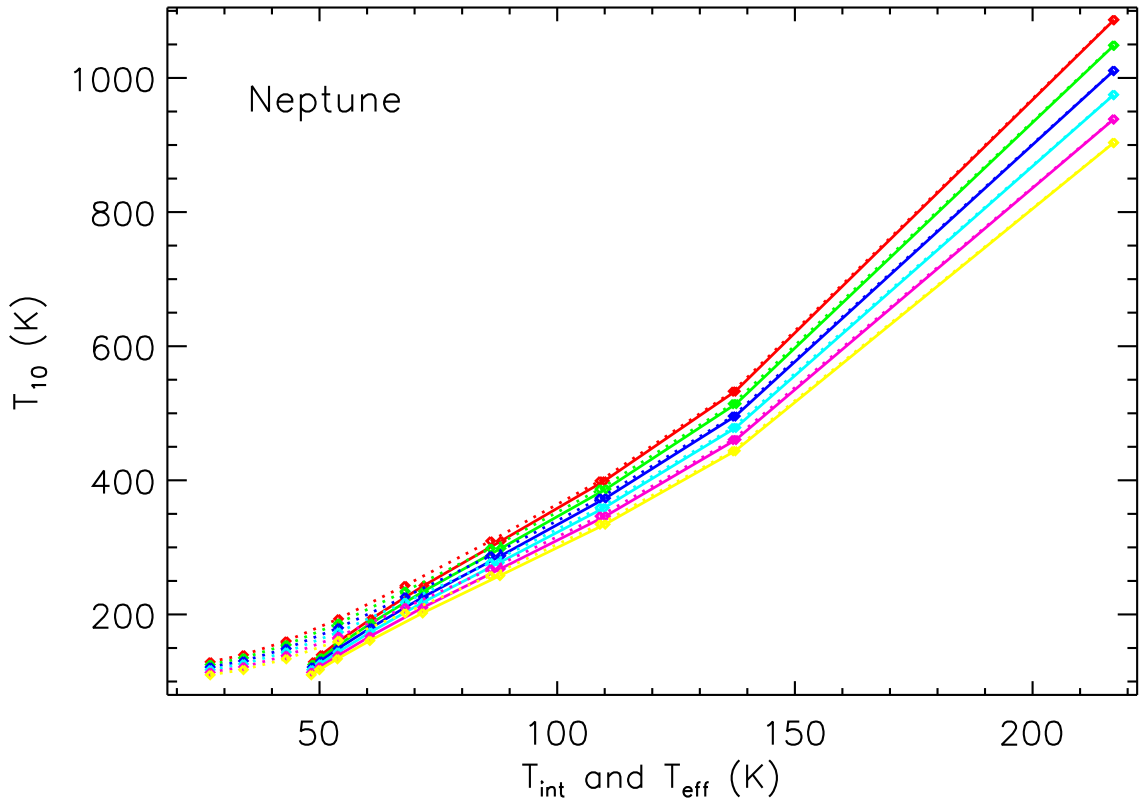}
\caption{The Neptune model atmosphere grid at 1.0 \ls.  Red is the lowest gravity, 4.0 m s$^{-2}$, while yellow is the highest, 12.6 m s$^{-2}$.  The grid is generally similar to that for Uranus, shown in Figure \ref{ugrid}, but since the incident flux upon the planet is smaller, one generally find a smaller \te\ for a given \ti.  The \te s are solid lines and \ti s are dotted lines.
\label{ngrid}}
\end{figure}

For Uranus and Neptune, we have computed four separate grids, at four different values of the incident flux.  The \ti\ ranges from 27 to 217 K, while the gravity range covers 0.45$\times$ Uranus' gravity to 1.1$\times$ Neptune's gravity.  Two values of the incident flux are for the current values  for Uranus and Neptune, at 19.2 and 30.1 AU from the current Sun, respectively.  The other two are at 1.8$\times$ the current flux received by Uranus, and 0.12$\times$ the current flux received by Neptune.  This very wide range allows for the inclusion of a time variable solar luminosity as well as the possibility of dramatic changes in the orbital distances of the planets with time \cp[e.g.][]{Gomes05}.  The grids at Uranus' and Neptune's current flux levels are compiled in Table 3 and shown in Figures \ref{ugrid} and \ref{ngrid}, respectively.  At high \te\ the two grids appear quite similar, but for Uranus' larger incident flux, the wider division between \ti\ and \te\, at low \ti, is readily apparent.  For nearly all models of the four planets, the atmosphere becomes convective and stays convective by 10 bars.  For a handful of models for Jupiter, the deep convective adiabat was not reached until $\sim$~15 bars, but the tabulated value of \tten\ is modified to be consistent with the entropy of the deep adiabat \cp[see][]{Burrows97}.

The model atmosphere grid, and its relation between $S$, \ti\ and \te, comes in through the energy conservation equation
\begin{equation} \label{eqL}
\frac{dL}{dm}= -T \frac{dS}{dt}
\end{equation}
where $L$ is the luminosity, $m$ is the mass variable, $T$ is the temperature in a mass shell, $S$ is the specific entropy of that shell, and $t$ is the time. After explicitly defining the dimensionless mass shell variable $m$ as
\begin{equation}
\label{shell}
m=\frac{1}{M}\int_{0}^{r} 4 \pi r^{{\prime}^2} dr^{\prime} \rho(r^{\prime}),
\end{equation}
we can rewrite Eq.~(\ref{eqL}) in terms of the time step $\partial t$, as
\begin{equation} \label{time}
\partial t = - \frac{M}{L} \int_{0}^{1}T \partial S dm,
\end{equation}
where $\partial t$ is the time step between two models that differ in entropy by $\partial S$.  The luminosity $L$ extracted from the planet is $4 \pi R^2 \sigma T_{\rm int}^4$, where the value of \ti\ at a given $S$ is given by the model atmospheres.  While at young ages \ti $\sim$ \te, at old ages, when \ti\ can be appreciably smaller than \te, interior cooling can be quite slow.   After presenting our new cooling calculations in the next section, in \S \ref{atmo} we will examine the affects of the new atmosphere grids, compared to previous tabulations.

\section{Cooling Calculations}
\subsection{Jupiter and Saturn}
Using these new model atmosphere grids, we can calculate the cooling history of the major planets.  Cooling calculations for these planets have been published by many authors.  Recent work for Jupiter and Saturn includes \ct{Hubbard99}, who explicitly showed the effects of a faint young Sun, and who included some limiting cases of additional interior energy due to the phase separation of helium from liquid metallic hydrogen at megabar pressures \cp{SS77b}.  \ct{FH03} looked at the evolution of Saturn and tested a number of previously published phase diagrams for H/He phase separation.  \ct{Saumon04} investigated the sensitivity of Jupiter cooling models to various hydrogen equations of state, which can predict quite different temperatures in the deep interior of the planet.  All of these models used the \ct{Burrows97} atmosphere grid.

In Figures \ref{js1} and \ref{js2} we compare Jupiter and Saturn evolution models to those of \ct{FH03}.  The old and new models make the same assumptions regarding solar luminosity and planetary structure, so the only differences are due the model atmospheres grids.  For both planets, a helium mass fraction ($Y$), relative to hydrogen, of 0.27 is used in the adiabatic H/He envelope. An rocky (zero-temperature ANEOS olivine) core mass of 10 \me\ is used for Jupiter, and 21 \me\ for Saturn.  Within the H/He envelope, the zero-temperature ANEOS water EOS is used to mix in a mass fraction of 0.059 and 0.030 of ``metals'' in Jupiter and Saturn, respectively.    In \ct{FH03} these choices reproduced the current radius and axial moment of inertia (obtained from more detailed structure models) at the time the planets cooled to their known \teff\ values.  As in \ct{FH03} the heat content of the rock and water are neglected in the evolution calculation.  \ct{FH03} use a Bond albedo 0.343 was assumed for both planets at all ages, while now we use the self-consistent grids.  The cooling times are modestly prolonged, as will be discussed in detail in \S \ref{atmo}.

A particular interesting difference shown in Figure \ref{js2} is the larger radii for the new planet models at young ages, which is most pronounced in Saturn.  This is a manifestation of the arbitrary initial condition (a hot, $\sim$~3 \rj, adiabatic sphere) along with the slowed cooling in the 700-400 K \ti\ range compared to previous models (See \S \ref{atmo}).  The initial conditions for cooling are tied to details of the energetics of planet formation, and are not well understood \cp{Marley07}.

The revised cooling age for Saturn does little to change the long-standing problem that the planet is much more luminous than homogeneous models predict \cp{Pollack77}.  This has long been attributed to the phase separation (``demixing'') of helium from liquid metallic hydrogen.  This helium solubility is thought to be minimized at pressures of several megabars, at pressures where the (gradual?) dissociation and ionization of hydrogen is nearly completed \cp{Stevenson75}.  The phase diagram of H/He mixtures is beyond the realm of current experiment, but recent advances in first principles calculations of the H/He phase diagram \cp{Lorenzen09,Morales09} should be tested in evolution models \cp{Hubbard99,FH03}, to see if the additional energy source from this ``helium rain'' can explain Saturn's luminosity.  Previously calculated phase diagrams were tested in \ct{FH03}, and were found to not explain Saturn's thermal evolution.  A complication that must be addressed in the future is whether the deep interior temperature gradient becomes dramatically super-adiabatic in the face of helium composition gradients \cp{SS77b,FH03}.

The model calculations for Jupiter, which yield an age of 5.3 Gyr, rather than 4.6 Gyr, could have important implications for the planet.  Jupiter's atmosphere is clearly depleted in helium, according to \emph{Galileo Entry Probe} data \cp{vonzahn98}, which is a strong indication that helium phase separation has occurred in this planet.  Furthermore the atmosphere's depletion in neon \cp{Mahaffy00} is strongly suggestive of helium demixing as well, as neon is expected to preferentially dissolve into helium-rich droplets \cp{Roulston95,Wilson10}.  The inclusion of the additional energy release due to this demixing will further prolong Jupiter's cooling \cp{Hubbard99}, leading to a worse match with observations.

However, \ct{Saumon04} have investigated cooling models of Jupiter with a variety of hydrogen EOSs, that predict a wide range of temperatures in Jupiter's deep interior, which led to evolutionary ages for homogeneous models as short as 3 Gyr. If the deep interior temperatures in Jupiter are lower than found with the \ct{SCVH} EOS used here, then it is possible that a combination of the model atmospheres presented here, helium rain (which prolongs the evolution), and colder interior temperatures (which quickens the evolution) could yield a good match to observations.  Recent work on the hydrogen EOS, both theoretically \cp{Nettelmann08,Militzer08}, and experimentally \cp{Holmes95,Collins01}, do yield  temperatures lower than predicted by \ct{SCVH}, so this avenue is plausible.

\subsection{Uranus and Neptune}
The long-standing problem for Uranus and Neptune has been that both of these planets are colder at the present day than cooling models predict.  This is a reverse of the situation for Saturn.  This issue is discussed in some detail in \ct{Ubook} and \ct{Nbook}, and in the general literature in \ct{Hubbard80b}.  In order to understand how advances in input physics over the past 30 years in the EOSs affect the thermal evolution of these planets, we have computed a set of evolution models that use the physics of the \ct{Hubbard80b} models, and compared them to our new calculations.  The models presented in \ct{Hubbard80b} have three distinct, adiabatic, layers.  The H/He envelope uses the EOS of \ct{Slattery76}, the ``icy'' layer uses a mixture the H$_2$O, NH$_3$, and CH$_4$ EOS from \ct{ZT78}, and the rocky core (a mixture of silicon, magnesium, iron, oxygen, and sulfur) is also from \ct{ZT78}.  \ct{Hubbard80b} also make estimates for the specific heat capacity of the icy and rocky layers.  Our implementation of the \ct{Hubbard80b} models agree well with the original, particularly for Uranus, and yield slightly shorter cooling times for both planets (\emph{dashed curves} in Fig.~\ref{unikoma}), compared to their work.

To investigate how updated EOSs affect the evolution, we can create cooling models that use the same ice-to-rock ratio as used in \ct{Hubbard80b}, 2.71-to-1, and the same \ct{Graboske75} atmosphere grid.  But instead we use updated EOSs for all three layers.  These are the \ct{SCVH} EOS for H/He and the Sesame EOSs of ``water 7154'' and ``dry sand'' \cp{SesameFort} for water and rock, respectively.  These tabulated ice and rock EOSs include calculations of the density- and temperature-dependent free energy at every temperature/density point, so no assumption must be made for the average heat capacity.  
In Figure \ref{unikoma} we compare cooling tracks with old and new EOSs.  Following previous work, we plot these tracks backwards in time from the current day, to see what initial values of \te\ could explain the current planets.  The updated EOS yield much faster evolution for both planets (\emph{dotted curves}).  The change for Neptune is enough to allow the planet have an initially ``hot start.''  However, Uranus models must start at a very low \te\ to explain the planet's current low \te.  Using these same new interior models, but with our new model atmosphere grids, yield the thick solid curves in Figure \ref{unikoma}.  These lead to slower cooling for both planets, most dramatically for Uranus.  Neptune's evolution is still approximately consistent with an arbitrarily hot start at formation.  This last finding agrees with the work published in \ct{Fortney09}.  

While the 3-layer models presented in \ct{Hubbard80b} were at the time consistent with observational constraints on their interior structure, that is no longer the case.  Neither the \ct{Hubbard80b} models, nor our new implementation of their 3-layer models (with the revised EOSs), are consistent with the gravity fields of these planets.  To further expand our treatment of Uranus and Neptune, we calculated new structure models, using the methods described in \ct{Nettelmann08} and \ct{Fortney09}.  We then investigated the thermal evolution of these models that are also consistent with the constraints on current structure.\footnote{These particular curves use \tone\ $=$ 73 and 78 K for the current Uranus and Neptune, to better allow the static structure model to match the \teff\ at the current time, at the expense of a slightly worse match to the observed \tone. However, changes in \tone\ by a few degrees in either direction at the current time, which adjusts the interior structure, has little effect on the cooling history.}

Like in \ct{Fortney09}, these models also use a three-layer structure, but include some water mixed into the H/He layer, and some H/He mixed into the water layer, above the rock core.  The outer layer is predominantly H/He with a helium mass fraction, relative to hydrogen ($Y$), of 0.27 \cp[the H-REOS and He-REOS equations of state][]{Nettelmann08}
with a 0.30 mass-fraction of water \cp[the EOS of][]{French09}.  The middle layer is mostly fluid water (mass fraction 0.878, beginning at 0.20 Mbar for Uranus, and 0.852, beginning at 0.10 Mbar for Neptune), with a small admixture of H/He.  Both of these layers are adiabatic.  The core is rock \cp{Hubbard89} with a mass of 1.51 \me\ for Uranus and 2.85 \me\ for Neptune.  For the evolution models, the rock core uses a radioactive luminosity from \ct{Guillot95} and a specific heat capacity of 1~Jg$^{-1}$K$^{-1}$.  The full allowed range of Uranus/Neptune interior compositions are explored in \ct{Fortney09}.  The fit to the \ct{Graboske75} grids use T$_1$=73 K ($K$=1.418) and T$_1$=78 K ($K$=1.571) for Uranus and Neptune, respectively, including the gravity dependence.

Evolutionary calculations for these model planets are shown in Figure \ref{un1}. We can reproduce well the current \teff\ of Neptune (blue models) with our new model atmosphere grid (\emph{solid curves}), as well as the older grid (\emph{dashed}). The latter models agree well with work published in \ct{Fortney09} which shows that the cooling times are insensitive to details of structure assumptions within the range allowed by the observational uncertainties.  Therefore, there is a plausible consistency for the planet:  the current interior structure and cooling history can be matched by one model.  As shown in Figure \ref{un1}, the mismatch for Uranus becomes appreciably larger with the new model atmospheres, in agreement with Figure \ref{unikoma}.  For Uranus in particular the model indicates a very slow change in \te\ with time in the current era, due to the larger incident flux and higher \teq\ than for Neptune.  Therefore small changes in the model atmosphere can lead to dramatic changes in cooling times, as seen in Figures \ref{unikoma} and \ref{un1}.  Another manifestation of Uranus's slow cooling is shown by the thin red curve in Figure \ref{unikoma}. This shows that a tiny change in the current \ti, to the maximum allowed by observation \cp[1$\sigma$ error bar,][]{Pearl91} can dramatically alter the calculation of the past cooling history.

\subsection{Effects of the Atmosphere Grids} \label{atmo}
\begin{figure} \epsscale{0.66}
\plotone{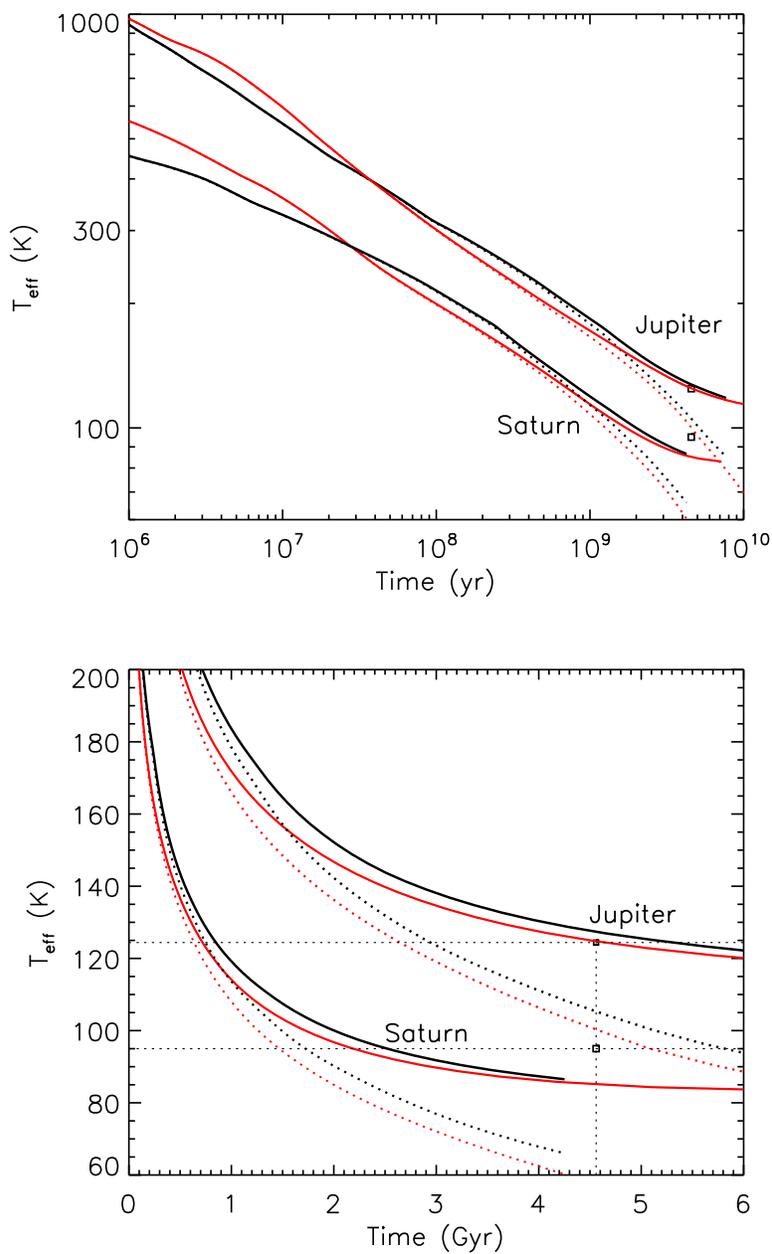}
\caption{Thermal evolution models of Jupiter and Saturn.  The upper panel shows the full evolutionary history, while the lower panel more clearly focuses on the current era.  Solid curves are \te, while dotted curves are \ti.  The thick black curves use the model atmosphere grids presented in this work, including the time-variable luminosity of the Sun, a linear increase from 0.72 \ls\ at time zero.  The thinner red curves utilize the same interior structure and solar luminosity, but instead use the \ct{Burrows97} grid \cp[see][]{FH03}.  The upper curves are for Jupiter while the lower curves are for Saturn.  The current uncertainty is the \te\ of each planet is represented by the size of the square at 4.56 Gyr in the lower panel.  The Jupiter model is over-luminous by 10\%, with a mean radius of 69,400 km.  The evolution of Saturn does not quite reach the current time, but a slight extrapolation off of the atmosphere grid shows the Saturn model has only 63\% of the actual planet's luminosity, with a mean radius 56,200 km. 
\label{js1}}
\end{figure}

\begin{figure} %\epsscale{0.66}
\plotone{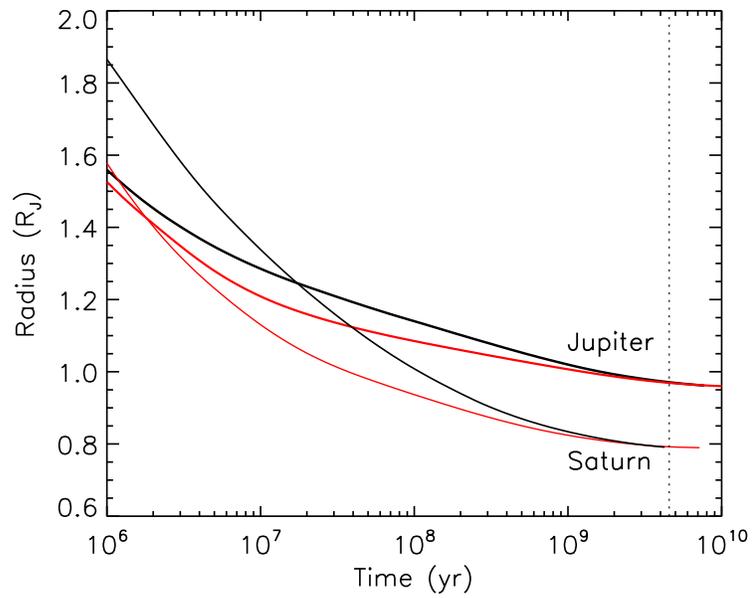}
\caption{Radius evolution of the models of Jupiter and Saturn shown in Figure \ref{js1} .  The black curves use the new model atmosphere grids presented in this work.  The red curves utilize the same interior structure and solar luminosity, but instead use the \ct{Burrows97} grid \cp[see][]{FH03}. 
\label{js2}}
\end{figure}

\begin{figure}\epsscale{1.0}
\plotone{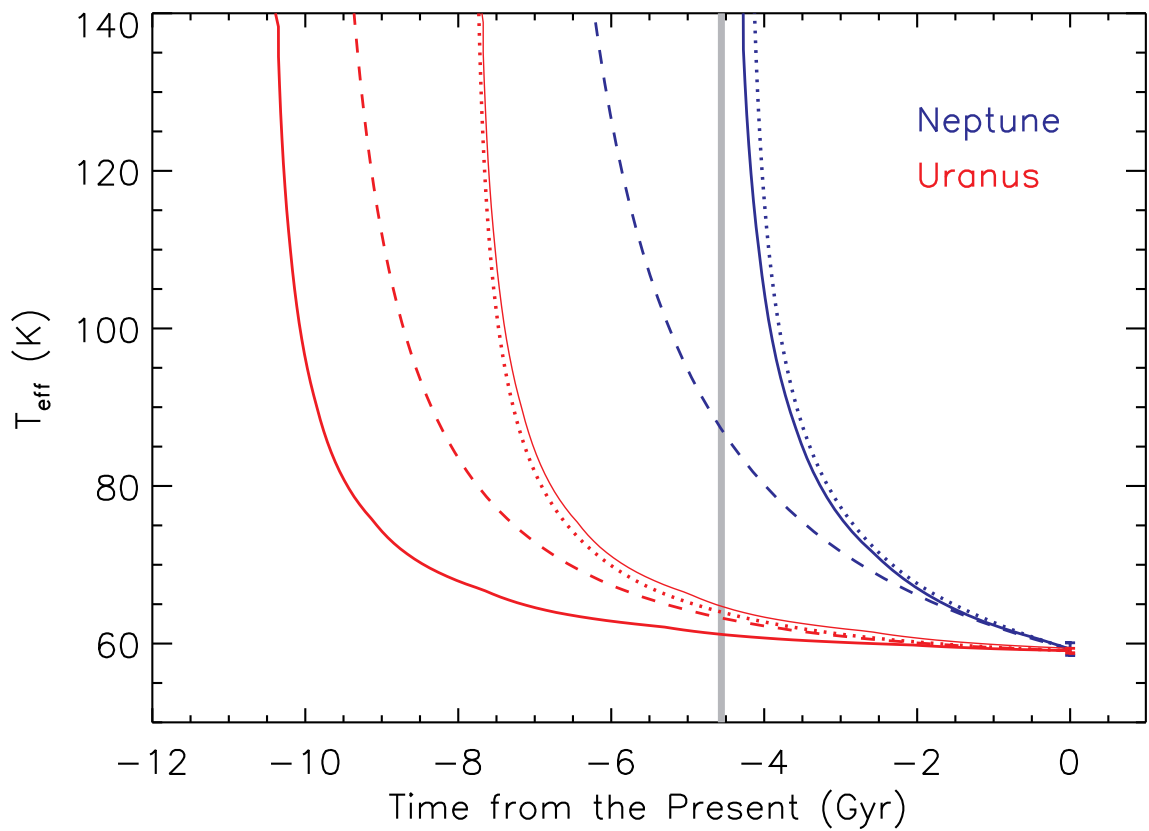}
\caption{Thermal evolution models of Uranus (red) and Neptune (blue) for models that feature three distinct homogeneous layers of H/He, fluid ices, and rock.  The models are constructed to match the \teff\ of both planets at the present day, and the evolution is followed backwards in time.  The thick grey line indicates the formation of the solar system.  The dashed curves are cooling calculations performed in the manner of \ct{Hubbard80b}, and use the \ct{Graboske75} model atmosphere grids.  The dotted curves use the same model atmospheres, but updated water and rock EOSs, which dramatically shorten the cooling times for both planets.  The thick solid curves use the same updated EOSs as the dotted-curve models, but also use our new model atmospheres.  The thin solid red curve, for Uranus, changes the thick-solid-red model only by adjusting the current \ti\ to the upper boundary of the 1$\sigma$ error bar.
\label{unikoma}}
\end{figure}

\begin{figure}
\plotone{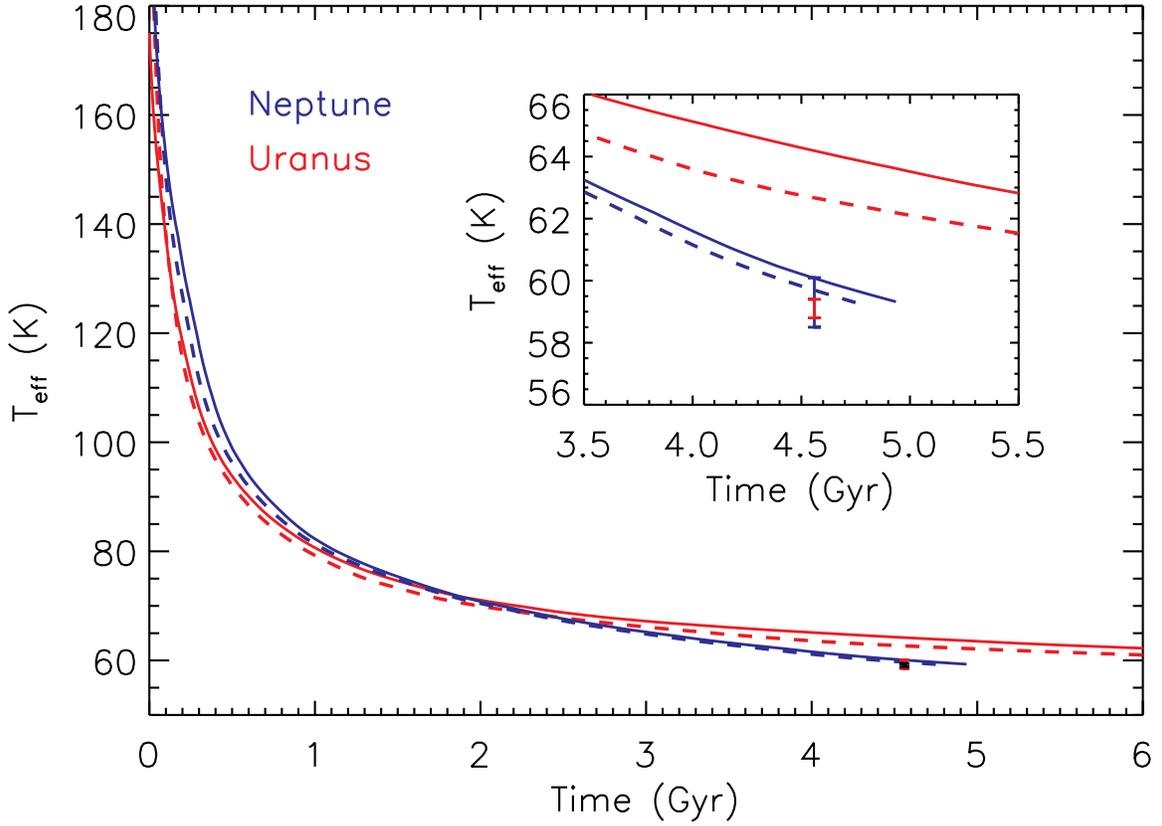}
\caption{Thermal evolution of Uranus (red) and Neptune (blue) for models that match the current gravity field constraints.  The thick solid curves use the model atmosphere grids presented in this work.  The dashed curves are similar to those presented in \ct{Fortney09}, which used a fit to the \ct{Graboske75} grid. When the \teff\ of each planet reaches the observed 
values, the interior structure of each planet is consistent with the gravity field.  At 4.56 Gyr, the radii of Uranus and Neptune are 25730 km and 24670 km, respectively.
\emph{Inset}: A zoom-in on the current era, which shows that the \teff\ of Neptune is well-reproduced, but the model Uranus is far too luminous.  Measured \teff\ and error bars are taken from \ct{Guillot05}.
\label{un1}}
\end{figure}

As shown in Figures \ref{js1}-\ref{un1}, a general finding of our work is that the cooling of our solar system's giant planets is slowed with the new model atmospheres.  The main reason is higher atmospheric opacity, due to improvements in opacity datbases, especially at high temperature, as well as the inclusion of opacity sources not previously known in 1975 or 1997. Here we investigate the reasons for these differences in atmospheric structure and cooling.  We will first examine Jupiter and Saturn.

It is important to note that the agreement between the \ct{Burrows97} grid and our grids are best at current (low) \ti\ values.  This is not necessarily surprising.  Both works use the same model atmosphere code \cp[that of][]{Marley96,MM99}.  However, the opacity databases used have changed significantly over the past 14 years, most importantly at high \ti\ values.  At low temperature, the only remaining opacity sources are methane vapor and H$_2$ collision induced absorption.  At warmer temperatures, observations of numerous T-type brown dwarfs have necessitated dramatic revisions to model atmospheres since 1997, including the inclusion of important new opacity sources, such as the alkali metals \cp{BMS,Allard01}.  In addition, high-temperature molecular opacity databases, which were in their infancy in 1997, are now becoming more complete, which generally leads to higher opacities.  Compared to Figure 2 of \ct{Burrows97} we find much smaller detached radiative zones below the photosphere, over a narrower range of \ti.  (In fact, for the Jupiter grid specifically, only the 450 K and 600 K models possess them.)

\begin{figure}
\plotone{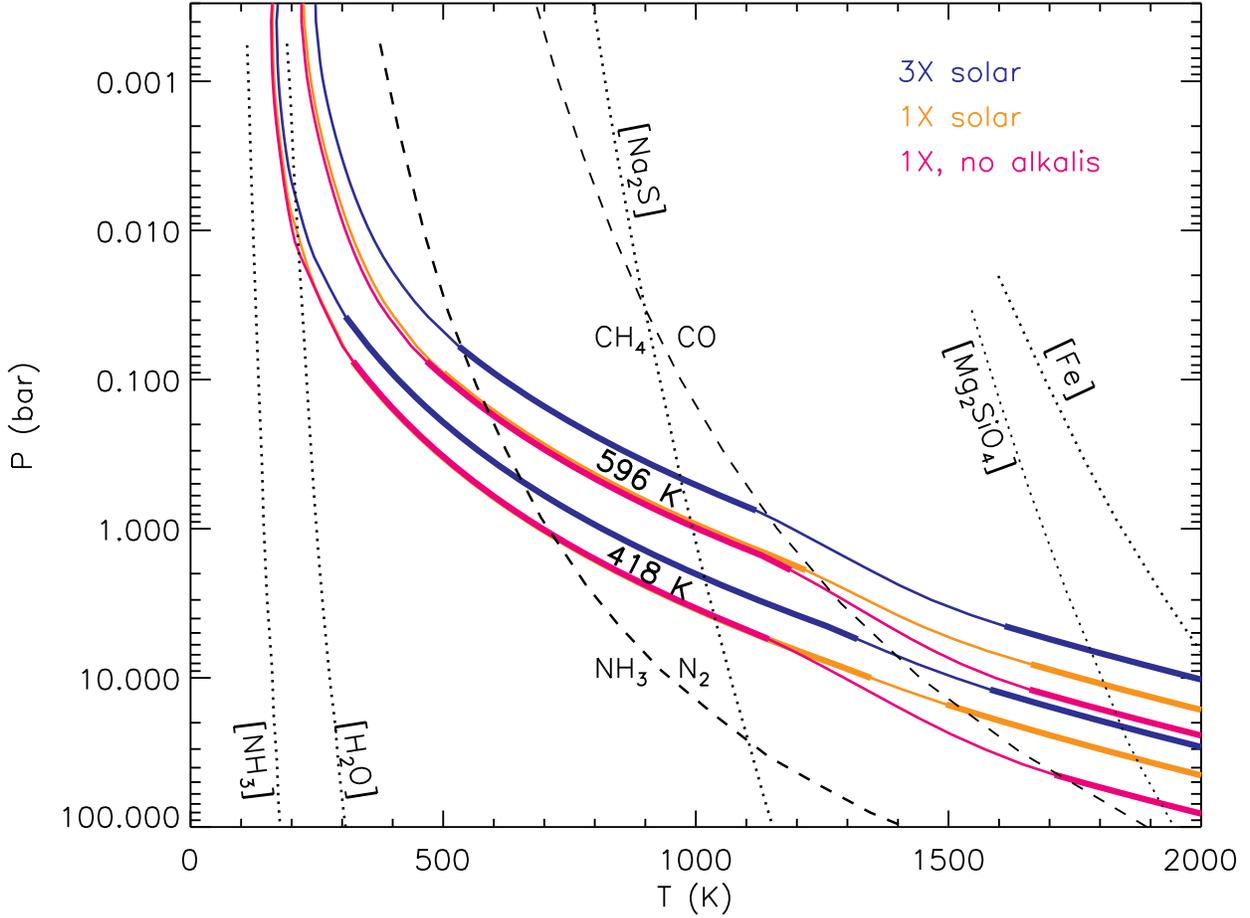}
\caption{Model pressure-temperature profiles of Jupiter at 417 K ($g=20.1$ m s$^{-2}$) and 596 K ($g=17.7$ m s$^{-2}$).  Thick lines indicate convective energy transport, while thin lines indiciate radiative energy transport.  Cloud condensation curves are in dotted black.  The equal-abundance curves of CH$_4$/CO and NH$_3$/N$_2$ are dashed.  The solid black curves represent the atmosphere of the Jupiter model at 32 and 10 Myr, and feature 3$\times$ solar metallicity.  The orange curves are the same, but feature solar metallicity.  The magenta curves are also solar metallicity, but with the opacity of neutral atomic Na and K atoms removed.  From black to orange, decreased metallicity leads to decreased opacity, a higher pressure photosphere, and lower specific entropy adiabat.  The removal of the alkali opacity leads to a detached radiative zone that is larger in extent with a shallower temperature gradient, again leading to a cooler adiabat.
\label{jintime2}}
\end{figure}

At contant \ti, a smaller or nonexistent detached radiative zone leads directly to a lower specific entropy adiabat, as shown in  Figure \ref{jintime2} for models at 596 and 417 K, representing Jupiter at ages of 10 and 32 Myr, respectively.  In black we show the standard 3$\times$ solar metallicity models.  In orange we show the same calculation, but with 1$\times$ solar metallicity.  The photospheric pressure (where $T$=\te) moves to higher pressure, due to the lower opacity, yielding a colder (lower specific entropy) adiabat.  The extent of the detached radiative zone is only marginally affected---the general trend towards smaller radiative zones at lower \te\ is not disturbed.  In magenta we plot the same 1$\times$ solar metallicity models, but with the Na and K alkali opacity removed, to show their affect.  The radiative zones are larger in vertical extent, with a shallower temperature gradient.  This shows that alkali opacity is a strong contributor towards closing the radiative zone, in a manner similar to that suggested by \ct{Guillot04} for Jupiter's current putative radiative zone at similar temperatures (1000-2000 K).  However, we note that even with alkali opacities removed, we are still unable to match the large extent of the radiative zones from Figure 2 of \ct{Burrows97}, which shows that overall molecular opacity updates \cp{Freedman08} since that time also contribute to the smaller radiative zones found today.

\begin{figure}
\plotone{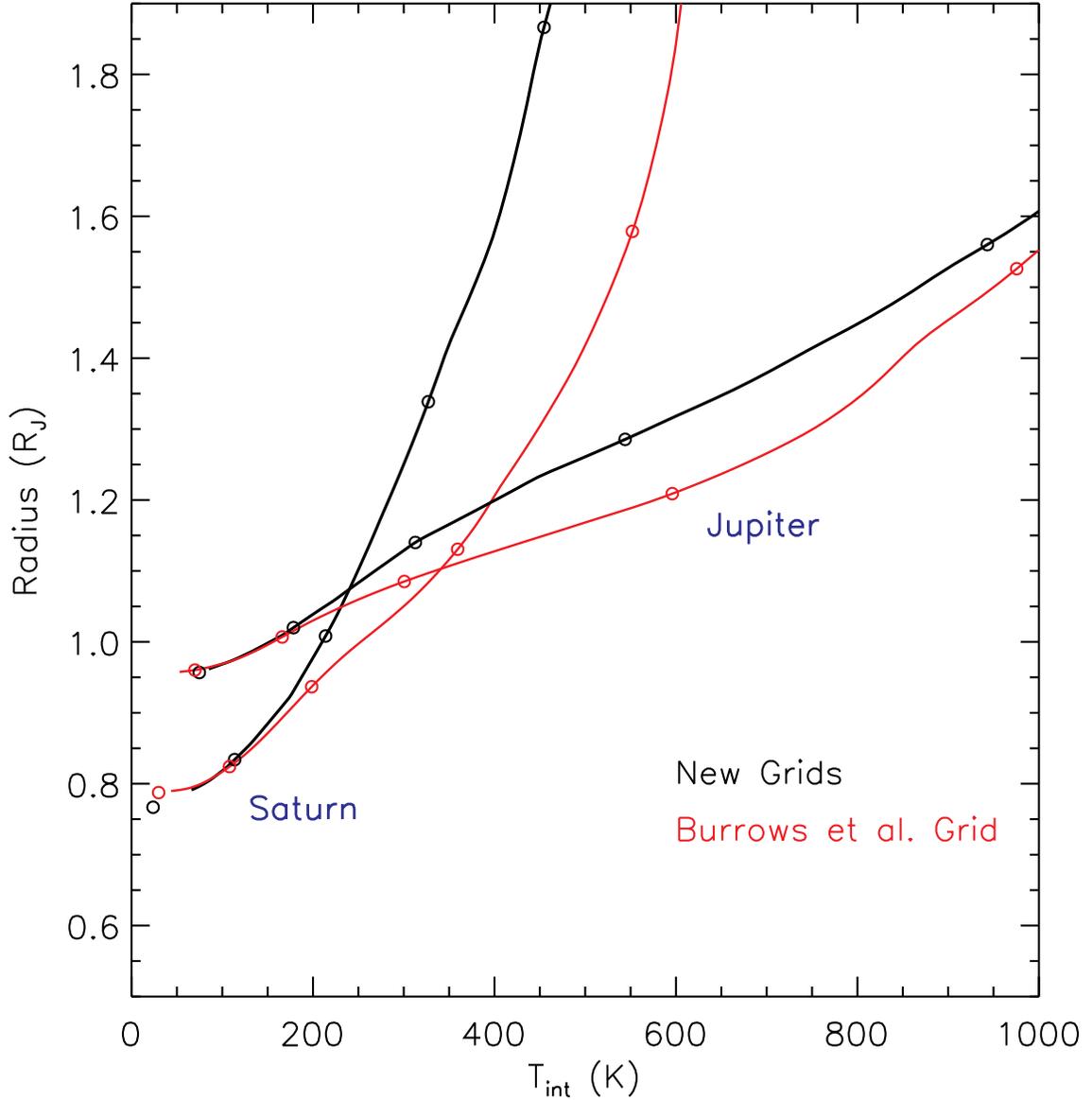}
\caption{Radii of models of Jupiter and Saturn from Figure \ref{js1} as a function of \ti. (1 \rj\ is 71492 km, Jupiter's equatorial radius.) Recall that radii, $S$, and \tten\ are all monotonic functions of each other.  As the planets cool, their radii contract, and $S$ and \tten\ decrease.  The open circles are time steps at ages of $10^6, 10^7, 10^8, 10^9$, and $10^{10}$ years.  Steeper $dR/dT_{\rm int}$ slopes lead to slower cooling.  (See text.)  For instance, a comparison with Figure \ref{js1} of \ti\ for Jupiter shows that the new (black) models, compared to the red models, cool more quickly at \ti\ $>800$ K, but more slowly below 800 K.  Below 200 K, the differences are quite small.  The same general picture is true for Saturn.
\label{jsgrid}}
\end{figure}

We now turn to the affect on evolution.  Given our discussion above, a given drop in $S$ between two identical structure models will occur at a lower \ti\ for a fully convective atmosphere, which leads to a longer time step in Equation \ref{time}, since this step goes at \ti\ to the fourth power.  Another way to look at cooling differences between the old and new atmosphere grids can be gleaned from Figure \ref{jsgrid}.  We plot planet radius vs.~\ti, as the radius can serve a proxy both for interior specific entropy (parameterized by \tten), and we can visually examine the surface gravity and \ti\ changes experienced in the model.  On this plot regions where the new atmosphere grid lead to slower cooling are where the slope $dR/dT_{\rm int}$, our proxy for $dS/dT_{\rm int}$, is steeper for the black curves than that of the red curves.  For a give change in \ti, they have a larger $\Delta S$, meaning a longer time step from Equation \ref{time}.  For example, a comparison with \ti\ vs.~time in Figure \ref{js1} shows that the new models lead to slower cooling from \ti\ of $\sim$800 K to the present \ti, although the differences are extremely small at \ti\ below 200 K.

\begin{figure}
\plotone{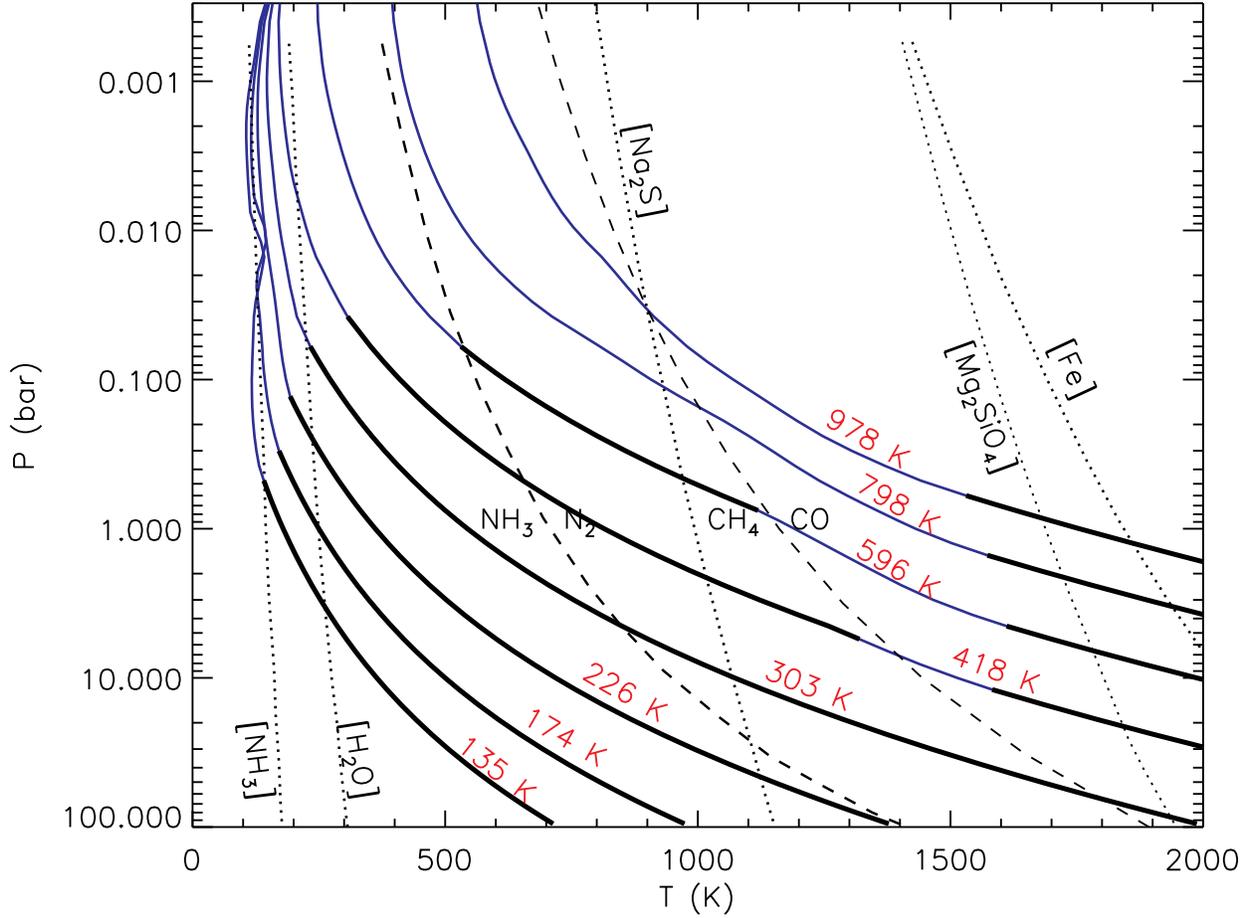}
\caption{Model pressure-temperature profiles of Jupiter from log $t$=6 to 9.5 years, in steps of 0.5.  The thin blue curves are radiative regions while thick black curves are convective regions.  Cloud condensation curves are in dotted black.  The equal-abundance curves of CH$_4$/CO and NH$_3$/N$_2$ are dashed.  The rightmost profile is \te $=976$ K, $g=11.4$ m s$^{-2}$ at 1 Myr.  The profiles 3rd and 4th from the right (596 K, 17.7 m s$^{-2}$, 10 Myr, and 417 K, 20.1 m s$^{-2}$, 32 Myr) have a detached convective zone.  From the models, it appears that Jupiter's water clouds formed at an age of $\sim$~30 Myr, while the ammonia clouds began to condense at $\sim$~1 Gyr.
\label{jintime}}
\end{figure}

In Figure \ref{jintime} we show model \emph{P--T} profiles for Jupiter's atmosphere, corresponding to the new cooling curve for Jupiter in Figure \ref{js1}.  While the early evolution is highly uncertain \cp[e.g.][]{Marley07}, we find that at 1 Myr Jupiter's atmosphere had only one (deep) convective zone, while a second, detached convective zone appears for a few tens of millions of years, from perhaps $\sim$5-50 Myr.  We also find that Jupiter's water clouds formed at an age of $\sim$~30 Myr, while the ammonia clouds began to condense at $\sim$~1 Gyr..  

\begin{figure}
\plotone{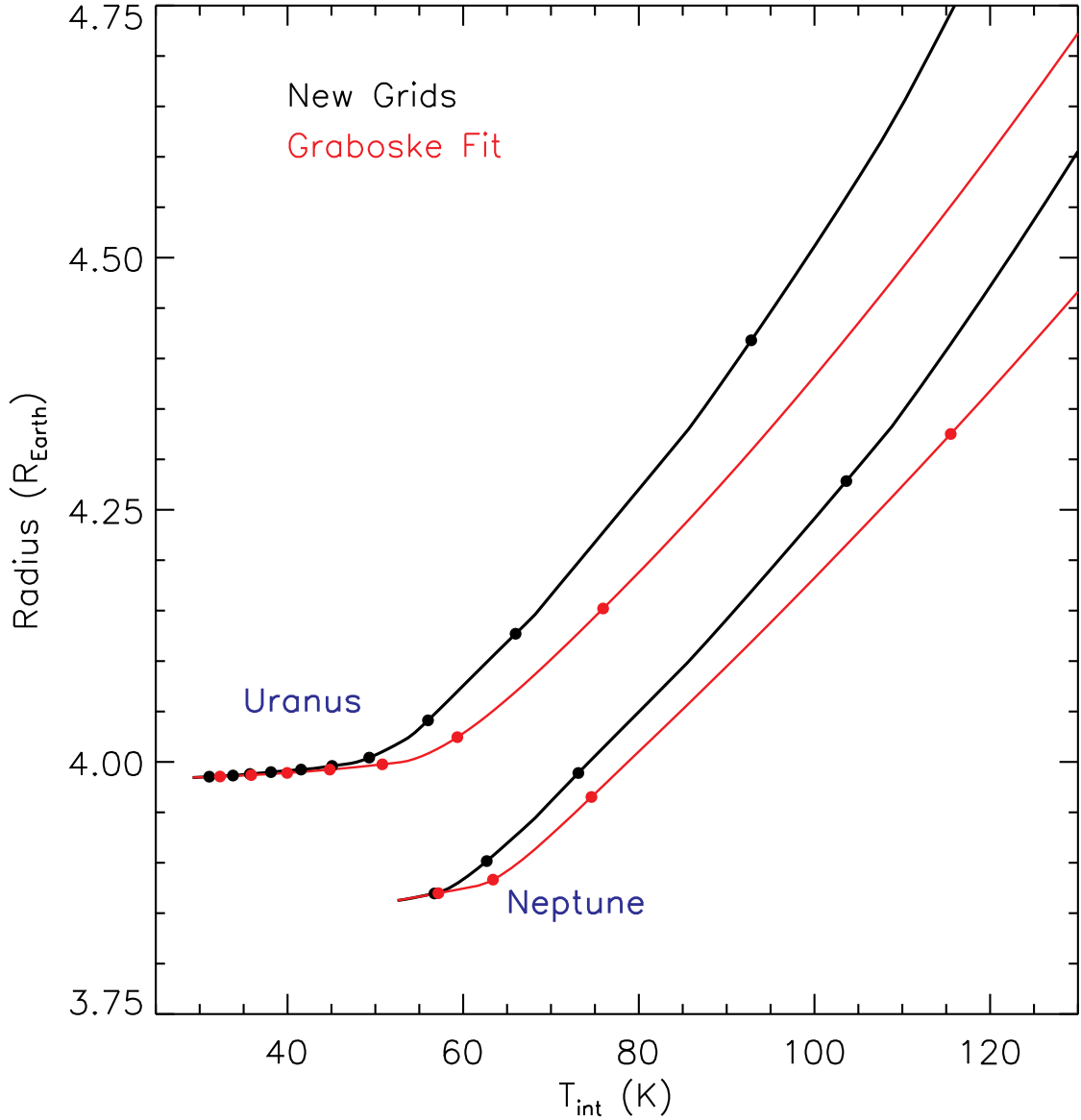}
\caption{Radii of models of Uranus and Neptune from Figure \ref{unikoma} as a function of \ti.  Recall that radii, $S$, and \tten\ all monotonic functions of each other.  Like Figure \ref{unikoma}, this plot is viewed as going backwards in time.  As the planets are warmer in the past, their radii enlarge, and $S$ and \tten\ increase.  The closed circles are time steps, equally space in 1 Gyr increments, since the initial models in the lower left.  This plot shows why cooling takes longer with the new grids (in black).  For Uranus, for instance, at 4.25 \re, a much smaller \ti\ indicates a longer time step $\partial t$ between successive models, as seen from Eq. \ref{time}.  Uranus, with its larger \teq, has much slower evolution in the past few gigayears, such that small changes in the atmosphere grids can lead to large changes in the evolution ages.
\label{ungrid}}
\end{figure}

Figure \ref{ungrid} shows that the Uranus and Neptune evolution models generally show an effect similar to that seen for Jupiter and Saturn.  Here the new grids always lead to a smaller \ti\ at a given radius.  Hence the cooling is always slower when using the grids, compared to those of \ct{Graboske75}.  It is clear from Figure \ref{ungrid} that the apparently small difference between the grids at low \ti\ becomes magnified because the cooling is so slow.  Even at higher \ti, farther back into the past, the differences between the two grids is larger than in Neptune.  There, the good agreement at low \ti\ is an effect of the \ct{Graboske75} grids being ad-hoc ``tuned'' to agree well with the current Uranus and Neptune, with $P_0$ reference pressure being changed from 1 bar to 0.750 bar for Uranus and 0.861 bar for Neptune \cp[see][]{Hubbard80b}.  As one moves to higher \ti, there is no reason to expect this tuning of the grid to hold, so the old and new grids diverge.

\section{Discussion and Conclusions}
Our own giant planets serve as our calibrators for the evolutionary theory used to understand the thermal evolution of extrasolar giant planets \cp[e.g.][]{Hubbard02,Guillot05,Fortney09,Baraffe10}.  As new equations of state come online, better understanding of interior energy transport develops, or new calculations of atmospheric structure or opacities are made, the problem of the cooling of the giant planets needs to be revisited.  We have investigated ``standard'' cooling models for these planets, where the planetary luminosity is predominantly due to the slow release of remnant formation energy, within adiabatic layers beneath the radiative atmosphere.

If our calculations are taken at face value, we can robustly conclude, as have authors before us, that Saturn is quite overluminous, necessitating a large role for helium phase separation over the past 2 Gyr \cp{SS77b,FH03}.  For Jupiter, the model cooling age is now modestly too long, which may suggest that the interior temperatures are colder than yielded by our models, as has been hinted at previously.  Although \ct{Hubbard99} have shown that cooling ages for Jupiter of 3.5 Gyr are possible in the absence of incident solar flux (a Bond albedo of 1) for this H/He EOS, extreme atmospheric reflectivity does not seem to be a realistic pathway to faster cooling.

Like others, we find that Uranus is quite underluminous.  \ct{Ubook} have suggested that a statically stable interior, unable to convect due to composition gradients, may be the best explanation.  However, these same models for Neptune can match the planet's \te\ at age of 4.5 Gyr.  The improvement is based predominantly on modern updates to the EOSs of constituent  materials.  It appears that whatever deep interior complications arise in Uranus do not arise in Neptune, or to a much smaller degree.  This could have profound implications for convection in these planets, and in particular the generation of these planets' non-axial non-dipolar-dominated magnetic fields \cp[see, e.g][for a review of dynamo modeling]{Stanley09}.

Using the interior statically stable geometry suggested by \ct{Ubook} and \ct{Nbook}, \ct{Stanley04,Stanley06} were able to reproduce the main features of the magnetic fields of Uranus and Neptune by hypothesizing dynamo action in only an outer shell in both planets.  However, we find that such a picture may not be viable for Neptune.  More recently \ct{Gomez07} have constructed 3D dynamo models with radial variable electrical conductivity, and found similar good agreement with observations without adhering to the statically stable interior geometry.  More work in this area is certainly needed.

\begin{figure}
\plotone{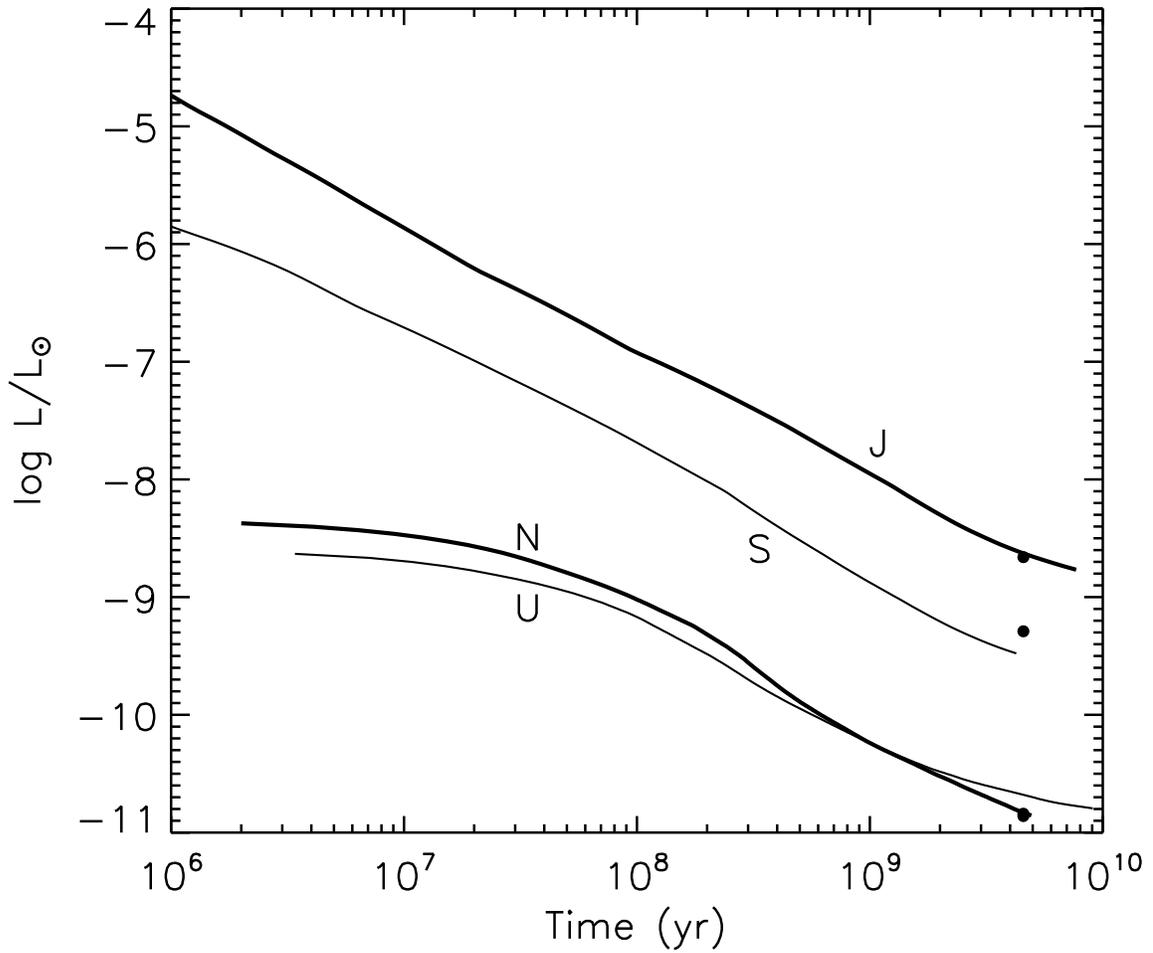}
\caption{Luminosity of the giant planets as a function of age.  Observed values \cp{Pearl91} are shown as black dots.  The values for Uranus and Neptune nearly overlap.  Estimates at young ages are highly uncertain, and depend strongly on the details of the formation process, especially for Uranus and Neptune.  Note that Uranus and Neptune cross at age of 1 Gyr, and Neptune becomes cooler.
\label{all}}
\end{figure}

The summary figure of the models presented here---the luminosity of the giant planets with age---is shown in Figure \ref{all}.  Of course the luminosities at young ages for all planets are uncertain, and depend strongly on the details of the formation process \cp{Marley07}.  Lower post-formation luminosities are certainly realistic, although at gigayear ages the details are not important.  In addition the cooling curve of Uranus should be regarded with suspicion.

One could certainly envision more complex models of the interiors of these planets, particularly for Uranus and Neptune.  We have modeled the ``fluid ice" component of these planets with a water EOS, but some previous authors have chosen instead to use mixtures of water, ammonia, and methane.  Shock experiments have studied a C-N-O-H mixture called ``synthetic Uranus" for use in ice giant modeling \cp{Nellis97,Hubbard91}.  The relative amounts of C, N, and O in these planet is not constrained by structure models or formation theory.  Also, one must remember that the high-pressure EOS of planetary material are uncertain.  \ct{Baraffe08} have quantitatively explored the use of different EOS for water, rock, and iron on the structure and evolution of Jupiter- and Neptune-class exoplanets.  The evolution of Uranus and Neptune could be most affected by EOS uncertainties.

We also caution that the good agreement with observations for the Neptune cooling models could be a coincidence.  Theory and experiment have probed the phase diagram of pure carbon, and have shown that it is solid diamond at Neptune-interior $P-T$ conditions.  A rain of solid diamond has been suggested for the interiors of Uranus and Neptune \cp{Ross81,Benedetti99,Eggert10}, which could, at least in principle, be an additional energy source that preferentially powers Neptune more strongly than Uranus, to explain their dichotomy.  These avenues should be explored in the future.

There are a few paths towards improving the atmosphere grids presented here.  For Jupiter and Saturn, one could investigate the reduced He/H ratio as the planets' age, which would affect the hydrogen collision induced absorption (CIA) that is an important infrared opacity source in these atmospheres.  We recommend updating CIA opacity in general, as the state-of-the-art in H-CIA opacity calculations \cp[e.g.][]{Borysow02} are now showing some mismatches in modeling brown dwarf spectra \cp{Cushing08}.  The inclusion of condensate clouds, either from equilibrium chemistry, such as ammonia and water, or a methane-derived photochemical haze, are important in accurately modeling the energy balance and temperature structure of these atmospheres.  Whether this could be understood well enough to predict with confidence the effects of clouds over the range of past \teff, atmospheric chemical mixing ratios, and surface gravity, in a 1D planet-wide average model atmosphere, is an open question.  It seems likely that further improvement towards understanding the cooling of the planets will likely come from work on the EOS of planetary materials.

\acknowledgements
J.~J.~F.~ acknowledges the support of NASA Outer Planets Research Program grant NNX08AU31G and the Alfred P.~Sloan Foundation.  M.~S.~M.~  gratefully acknowledges the influence of Jim Pollack, who was in the planning phases of a similar study at the time of his untimely passing.  Our atmospheric modeling is strongly influenced by his work.  Referee Gilles Chabrier provided valuable comments that improved the draft.   

%\bibliographystyle{apj}
%\bibliography{references}

\input{jgrids}

\input{sgrids}
\input{ungrids}

\end{document}

%% file: jgrids.tex
\begin{center}
\begin{deluxetable}{cccccc}
\tablecolumns{6}
\tablewidth{0pc}
\tabletypesize{\footnotesize}
\tablecaption{Jupiter grid at 1 and 0.7 \ls}
\tablehead{
\colhead{gravity}  &  \colhead{\te, 1.0}  &  \colhead{\tten, 1.0} &  \colhead{\te, 0.7}  &  \colhead{\tten, 0.7} &  \colhead{\ti}}
\startdata 
3.2	&	1201.6	&	3471.1	&	1201.6	&	3471.1	&	1200	\\
	&	904.7	&	3124.9	&	904.7	&	3124.9	&	900	\\
	&	751.3	&	2894.2	&	751.2	&	2894.2	&	750	\\
	&	600.6	&	2582.3	&	600.5	&	2582.4	&	600	\\
	&	-	&	-	&	-	&	-	&	450	\\
	&	-	&	-	&	-	&	-	&	316	\\
	&	-	&	-	&	-	&	-	&	224	\\
	&	-	&	-	&	-	&	-	&	168	\\
	&	-	&	-	&	-	&	-	&	133	\\
	&	-	&	-	&	-	&	-	&	112	\\
	&	-	&	-	&	-	&	-	&	100	\\
	&	-	&	-	&	-	&	-	&	89	\\
5.6	&	1201.2	&	3359.3	&	1202.6	&	3360.3	&	1200	\\
	&	902.5	&	2984.3	&	902.6	&	2984.3	&	900	\\
	&	750.8	&	2730.3	&	750.7	&	2730.2	&	750	\\
	&	600.4	&	2391.7	&	600.4	&	2391.6	&	600	\\
	&	450.6	&	1966.2	&	450.4	&	1966.1	&	450	\\
	&	-	&	-	&	-	&	-	&	316	\\
	&	-	&	-	&	-	&	-	&	224	\\
	&	-	&	-	&	-	&	-	&	168	\\
	&	-	&	-	&	-	&	-	&	133	\\
	&	-	&	-	&	-	&	-	&	112	\\
	&	-	&	-	&	-	&	-	&	100	\\
	&	-	&	-	&	-	&	-	&	89	\\
9.1	&	1201.8	&	3255.1	&	1202.2	&	3255.8	&	1200	\\
	&	901.5	&	2850.3	&	901.5	&	2850.2	&	900	\\
	&	750.5	&	2571.4	&	750.5	&	2571.4	&	750	\\
	&	600.4	&	2222.1	&	600.3	&	2222.1	&	600	\\
	&	450.6	&	1816.0	&	450.4	&	1815.9	&	450	\\
	&	317.4	&	1312.2	&	317.0	&	1310.2	&	316	\\
	&	227.3	&	842.6	&	226.4	&	838.0	&	224	\\
	&	-	&	-	&	-	&	-	&	168	\\
	&	-	&	-	&	-	&	-	&	133	\\
	&	-	&	-	&	-	&	-	&	112	\\
	&	-	&	-	&	-	&	-	&	100	\\
	&	-	&	-	&	-	&	-	&	89	\\
13.5	&	1201.5	&	3165.6	&	1201.8	&	3166.0	&	1200	\\
	&	901.5	&	2731.4	&	901.0	&	2731.4	&	900	\\
	&	750.4	&	2435.8	&	750.4	&	2435.8	&	750	\\
	&	600.3	&	2086.1	&	600.3	&	2086.0	&	600	\\
	&	450.6	&	1701.6	&	450.4	&	1701.6	&	450	\\
	&	317.4	&	1233.5	&	317.0	&	1231.7	&	316	\\
	&	227.4	&	791.8	&	226.4	&	787.4	&	224	\\
	&	175.3	&	558.7	&	173.2	&	552.4	&	168	\\
	&	146.4	&	451.4	&	142.7	&	439.9	&	133	\\
	&	-	&	-	&	-	&	-	&	112	\\
	&	-	&	-	&	-	&	-	&	100	\\
	&	-	&	-	&	-	&	-	&	89	\\
18.2	&	-	&	-	&	-	&	-	&	1200	\\
	&	-	&	-	&	-	&	-	&	900	\\
	&	750.3	&	2330.5	&	750.3	&	2330.5	&	750	\\
	&	600.3	&	1987.6	&	600.2	&	1987.6	&	600	\\
	&	450.6	&	1617.7	&	450.4	&	1617.6	&	450	\\
	&	317.4	&	1174.5	&	317.0	&	1172.6	&	316	\\
	&	227.4	&	754.4	&	226.4	&	750.4	&	224	\\
	&	175.4	&	533.5	&	173.2	&	527.5	&	168	\\
	&	146.4	&	430.9	&	142.8	&	419.6	&	133	\\
	&	131.9	&	375.8	&	126.8	&	360.0	&	112	\\
	&	125.0	&	348.9	&	118.9	&	330.1	&	100	\\
	&	119.7	&	328.0	&	112.9	&	306.9	&	89	\\
22.4	&	-	&	-	&	-	&	-	&	1200	\\
	&	-	&	-	&	-	&	-	&	900	\\
	&	-	&	-	&	-	&	-	&	750	\\
	&	600.3	&	1918.6	&	600.2	&	1918.6	&	600	\\
	&	450.6	&	1561.0	&	450.4	&	1560.9	&	450	\\
	&	317.4	&	1135.3	&	317.0	&	1133.5	&	316	\\
	&	227.4	&	729.5	&	226.4	&	725.3	&	224	\\
	&	175.4	&	515.9	&	173.3	&	510.0	&	168	\\
	&	146.5	&	417.1	&	142.8	&	406.5	&	133	\\
	&	131.9	&	364.3	&	126.8	&	348.5	&	112	\\
	&	125.0	&	337.9	&	119.0	&	320.0	&	100	\\
	&	119.8	&	317.9	&	113.0	&	296.9	&	89	\\
\tablebreak
25.1	&	-	&	-	&	-	&	-	&	1200	\\
	&	-	&	-	&	-	&	-	&	900	\\
	&	-	&	-	&	-	&	-	&	750	\\
	&	-	&	-	&	-	&	-	&	600	\\
	&	-	&	-	&	-	&	-	&	450	\\
	&	-	&	-	&	-	&	-	&	316	\\
	&	227.4	&	716.0	&	226.4	&	712.0	&	224	\\
	&	175.4	&	506.9	&	173.3	&	501.1	&	168	\\
	&	146.5	&	409.7	&	142.8	&	399.2	&	133	\\
	&	132.0	&	357.4	&	126.9	&	342.3	&	112	\\
	&	125.1	&	331.9	&	119.1	&	314.3	&	100	\\
	&	119.9	&	312.5	&	113.1	&	292.3	&	89	\\
28.2	&	-	&	-	&	-	&	-	&	1200	\\
	&	-	&	-	&	-	&	-	&	900	\\
	&	-	&	-	&	-	&	-	&	750	\\
	&	-	&	-	&	-	&	-	&	600	\\
	&	-	&	-	&	-	&	-	&	450	\\
	&	-	&	-	&	-	&	-	&	316	\\
	&	227.7	&	703.7	&	226.6	&	699.5	&	224	\\
	&	176.0	&	499.4	&	173.7	&	493.1	&	168	\\
	&	146.5	&	402.3	&	142.8	&	391.6	&	133	\\
	&	132.0	&	350.9	&	126.9	&	336.1	&	112	\\
	&	125.1	&	325.8	&	119.1	&	308.7	&	100	\\
	&	120.0	&	307.2	&	113.2	&	286.8	&	89	\\
\enddata
\tablecomments{Surface gravities are in m s$^{-2}$.  A metallicity of [M/H]=+0.5 ($\sim$~3$\times$ solar) is assumed.  ``0.7'' and ``1.0'' mean 0.7 and 1.0 times the current Jovian incident flux.}
\label{jtab}
\end{deluxetable}
\end{center}

%% file: sgrids.tex
\begin{center}
\begin{deluxetable}{cccccc}
\tablecolumns{6}
\tablewidth{0pc}
\tabletypesize{\footnotesize}
\tablecaption{Saturn grids at 1 and 0.7 \ls}
\tablehead{
\colhead{gravity}  &  \colhead{\te, 1.0}  &  \colhead{\tten, 1.0} &  \colhead{\te, 0.7}  &  \colhead{\tten, 0.7} &  \colhead{\ti}}
\startdata 
1.3	&		652.8	&		3196.1	&		652.8	&		3196.1	&	650\\
	&		551.5	&		3020.1	&		551.4	&		3020.1	&	550\\
	&		450.7	&		2750.7	&		450.7	&		2750.6	&	450\\
	&		350.5	&		2232.2	&		350.4	&		2231.6	&	350\\
	&	-		&	-		&	-		&	-		&	275\\
	&	-		&	-		&	-		&	-		&	225\\
	&	-		&	-		&	-		&	-		&	175\\
	&	-		&	-		&	-		&	-		&	131\\
	&	-		&	-		&	-		&	-		&	104\\
	&	-		&	-		&	-		&	-		&	87\\
	&	-		&	-		&	-		&	-		&	78\\
	&	-		&	-		&	-		&	-		&	69\\
3.0	&		651.3	&		2993.9	&		651.3	&		2993.8	&	650\\
	&		550.7	&		2783.9	&		550.7	&		2783.9	&	550\\
	&		450.4	&		2485.2	&		450.4	&		2485.2	&	450\\
	&		350.3	&		2000.7	&		350.3	&		2000.1	&	350\\
	&		275.5	&		1500.1	&		275.6	&		1498.8	&	275\\
	&		226.1	&		1120.5	&		225.8	&		1118.4	&	225\\
	&	-		&	-		&	-		&	-		&	175\\
	&	-		&	-		&	-		&	-		&	131\\
	&	-		&	-		&	-		&	-		&	104\\
	&	-		&	-		&	-		&	-		&	87\\
	&	-		&	-		&	-		&	-		&	78\\
	&	-		&	-		&	-		&	-		&	69\\
5.4	&		650.7	&		2831.4	&		650.7	&		2831.4	&	650\\
	&		550.4	&		2597.4	&		550.4	&		2597.4	&	550\\
	&		450.3	&		2285.2	&		450.2	&		2285.1	&	450\\
	&		350.4	&		1844.9	&		350.3	&		1844.4	&	350\\
	&		275.6	&		1376.2	&		275.5	&		1375.0	&	275\\
	&		226.1	&		1021.8	&		225.8	&		1019.7	&	225\\
	&		177.2	&		692.2	&		176.5	&		689.2	&	175\\
	&		135.7	&		490.5	&		134.3	&		484.9	&	131\\
	&		112.2	&		380.9	&		109.9	&		373.3	&	104\\
	&	-		&	-		&	-		&	-		&	87\\
	&	-		&	-		&	-		&	-		&	78\\
	&	-		&	-		&	-		&	-		&	69\\
7.2	&		-	&		-	&		-	&		-	&	650\\
	&		-	&		-	&		-	&		-	&	550\\
	&		-	&		-	&		-	&		-	&	450\\
	&		350.4	&		1769.9	&		350.3	&		1769.4	&	350\\
	&		275.6	&		1318.1	&		275.5	&		1317.0	&	275\\
	&		226.1	&		976.6	&		225.7	&		974.6	&	225\\
	&		177.1	&		661.7	&		176.6	&		658.7	&	175\\
	&		135.7	&		478.8	&		134.3	&		463.4	&	131\\
	&		112.2	&		364.8	&		110.0	&		356.5	&	104\\
	&		99.6	&		311.5	&		96.0	&		302.4	&	87\\
	&	-		&	-		&	-		&	-		&	78\\
	&	-		&	-		&	-		&	-		&	69\\
8.9	&	-		&	-		&	-		&	-		&	650\\
	&	-		&	-		&	-		&	-		&	550\\
	&	-		&	-		&	-		&	-		&	450\\
	&	-		&	-		&	-		&	-		&	350\\
	&	-		&	-		&	-		&	-		&	275\\
	&		226.1	&		944.6	&		225.8	&		942.4	&	225\\
	&		177.2	&		640.2	&		176.6	&		637.9	&	175\\
	&		135.8	&		453.6	&		134.3	&		448.4	&	131\\
	&		112.3	&		353.3	&		110.0	&		345.1	&	104\\
	&		99.5	&		301.0	&		96.4	&		291.7	&	87\\
	&		93.9	&		279.2	&		89.9	&		268.2	&	78\\
	&		89.2	&		261.5	&		84.1	&		246.3	&	69\\
10.0	&	-		&	-		&	-		&	-		&	650\\
	&	-		&	-		&	-		&	-		&	550\\
	&	-		&	-		&	-		&	-		&	450\\
	&	-		&	-		&	-		&	-		&	350\\
	&	-		&	-		&	-		&	-		&	275\\
	&	-		&	-		&	-		&	-		&	225\\
	&		177.2	&		628.6	&		176.6	&		625.7	&	175\\
	&		135.8	&		445.4	&		134.4	&		440.0	&	131\\
	&		112.3	&		347.5	&		110.0	&		339.2	&	104\\
	&		99.5	&		296.9	&		96.1	&		286.5	&	87\\
	&		94.0	&		275.9	&		90.3	&		264.2	&	78\\
	&		89.2	&		257.0	&		84.1	&		242.3	&	69\\
11.2	&	-		&	-		&	-		&	-		&	650\\
	&	-		&	-		&	-		&	-		&	550\\
	&	-		&	-		&	-		&	-		&	450\\
	&	-		&	-		&	-		&	-		&	350\\
	&	-		&	-		&	-		&	-		&	275\\
	&	-		&	-		&	-		&	-		&	225\\
	&	-	177.2	&		617.4	&	-	176.6	&	-	614.9	&	175\\
	&		135.7	&		437.1	&		134.4	&		432.0	&	131\\
	&		112.4	&		341.7	&		110.2	&		333.3	&	104\\
	&		99.5	&		291.6	&		96.1	&		282.4	&	87\\
	&		94.0	&		270.3	&		90.1	&		259.9	&	78\\
	&		89.2	&		252.8	&		84.2	&		238.3	&	69\\
12.0	&	-		&	-		&	-		&	-		&	650\\
	&	-		&	-		&	-		&	-		&	550\\
	&	-		&	-		&	-		&	-		&	450\\
	&	-		&	-		&	-		&	-		&	350\\
	&	-		&	-		&	-		&	-		&	275\\
	&	-		&	-		&	-		&	-		&	225\\
	&	-		&	-		&	-		&	-		&	175\\
	&		135.8	&		432.3	&		134.4	&		427.4	&	131\\
	&		112.5	&		338.1	&		110.2	&		329.7	&	104\\
	&		99.6	&		288.3	&		96.2	&		279.2	&	87\\
	&		94.0	&		267.3	&		90.1	&		257.1	&	78\\
	&		89.2	&		250.2	&		84.2	&		235.9	&	69\\
	\enddata
\tablecomments{Surface gravities are in m s$^{-2}$.  A metallicity of [M/H]=+1.0 (10$\times$ solar) is assumed.  ``0.7'' and ``1.0'' mean 0.7 and 1.0 times the current Saturnian incident flux.}
\label{stab}
\end{deluxetable}
\end{center}

%% file: ungrids.tex
\begin{center}
\begin{deluxetable}{cccccccccccccccccc}
\rotate
\tablecolumns{14}
\tablewidth{0pc}
\tabletypesize{\footnotesize}
\tablecaption{Uranus and Neptune Grids at 4 Level of Solar Flux}
\tablehead{
\colhead{gravity}  &  \colhead{\te, 0.12N}  &  \colhead{\tten, 0.12N}&  \colhead{\tone, 0.12N}  &  \colhead{\te, 1.0N}  &  \colhead{\tten, 1.0N} &  \colhead{\tone, 1.0N}&  \colhead{\te, 1.0U}  &  \colhead{\tten, 1.0U} &  \colhead{\tone, 1.0U}&  \colhead{\te, 1.8U}  &  \colhead{\tten, 1.8U} &  \colhead{\tone, 1.8U}&    \colhead{\ti}}
\startdata 
4.0	&	32.66	&	88.64	&	49.11	&	48.71	&	129.32	&	66.00	&	59.87	&	164.80	&	80.55	&	69.54	&	196.07	&	94.22	&	27	\\
	&	37.37	&	105.39	&	56.31	&	50.38	&	139.77	&	70.22	&	60.80	&	171.55	&	83.42	&	70.14	&	200.74	&	96.34	&	34	\\
	&	44.89	&	134.43	&	68.06	&	54.11	&	160.32	&	78.67	&	63.04	&	186.10	&	89.75	&	71.63	&	211.46	&	101.29	&	43	\\
	&	55.11	&	177.20	&	85.86	&	60.87	&	193.49	&	93.05	&	67.67	&	211.78	&	101.44	&	74.30	&	229.80	&	109.99	&	54	\\
	&	68.46	&	231.89	&	111.00	&	71.96	&	243.06	&	116.47	&	76.39	&	253.96	&	121.88	&	81.78	&	267.16	&	128.53	&	68	\\
	&	86.23	&	301.80	&	146.36	&	88.21	&	309.59	&	150.42	&	90.90	&	315.21	&	153.36	&	94.32	&	321.08	&	156.43	&	86	\\
	&	109.12	&	396.53	&	196.11	&	109.93	&	399.17	&	197.50	&	111.22	&	404.77	&	200.44	&	113.40	&	409.02	&	202.66	&	109	\\
	&	137.31	&	531.03	&	265.95	&	137.48	&	532.84	&	266.87	&	138.41	&	535.66	&	268.32	&	139.49	&	539.71	&	270.40	&	137	\\
	&	217.06	&	1086.00	&	551.01	&	217.15	&	1086.72	&	551.39	&	217.22	&	1087.84	&	551.98	&	217.63	&	1090.67	&	553.46	&	217	\\
5.0	&	32.63	&	85.80	&	47.82	&	48.59	&	125.33	&	64.40	&	59.69	&	159.82	&	78.46	&	69.34	&	190.39	&	91.66	&	27	\\
	&	37.35	&	101.83	&	54.82	&	50.28	&	135.23	&	68.38	&	60.63	&	166.29	&	81.18	&	69.94	&	194.96	&	93.71	&	34	\\
	&	44.87	&	129.67	&	66.14	&	54.03	&	154.67	&	76.32	&	62.88	&	179.89	&	87.03	&	71.44	&	204.83	&	98.21	&	43	\\
	&	55.09	&	170.88	&	83.13	&	60.81	&	186.84	&	90.08	&	67.53	&	204.74	&	98.17	&	74.25	&	222.41	&	106.46	&	54	\\
	&	68.46	&	224.03	&	107.23	&	71.88	&	234.56	&	112.30	&	76.29	&	245.17	&	117.51	&	81.63	&	258.32	&	124.06	&	68	\\
	&	86.23	&	291.63	&	141.08	&	88.18	&	298.61	&	144.70	&	91.04	&	302.92	&	147.98	&	94.31	&	310.17	&	150.72	&	86	\\
	&	109.12	&	383.20	&	189.11	&	110.26	&	386.88	&	191.04	&	111.22	&	391.57	&	193.51	&	113.02	&	396.00	&	195.83	&	109	\\
	&	137.31	&	512.66	&	256.50	&	137.66	&	514.23	&	257.31	&	138.17	&	517.05	&	258.75	&	139.18	&	520.87	&	260.72	&	137	\\
	&	217.05	&	1048.68	&	531.44	&	217.13	&	1048.72	&	531.48	&	217.27	&	1049.94	&	532.10	&	217.64	&	1052.41	&	533.39	&	217	\\
6.3	&	32.60	&	83.23	&	46.61	&	48.47	&	121.55	&	62.88	&	59.53	&	154.87	&	76.40	&	69.15	&	184.65	&	89.11	&	27	\\
	&	37.32	&	98.37	&	53.36	&	50.21	&	130.57	&	66.51	&	60.48	&	160.84	&	78.89	&	69.76	&	188.84	&	90.97	&	34	\\
	&	44.85	&	124.96	&	64.25	&	53.95	&	149.20	&	74.06	&	62.74	&	173.88	&	84.43	&	71.26	&	198.27	&	95.22	&	43	\\
	&	55.08	&	164.66	&	80.49	&	60.75	&	180.04	&	87.10	&	67.41	&	197.53	&	94.88	&	74.20	&	215.52	&	103.18	&	54	\\
	&	68.70	&	217.54	&	104.15	&	71.83	&	226.18	&	108.26	&	76.19	&	236.65	&	113.32	&	81.54	&	249.39	&	119.60	&	68	\\
	&	86.39	&	282.31	&	136.28	&	88.14	&	287.47	&	138.93	&	90.73	&	293.36	&	141.97	&	94.31	&	301.78	&	146.35	&	86	\\
	&	109.12	&	369.37	&	181.83	&	110.17	&	373.36	&	183.93	&	111.20	&	377.69	&	186.21	&	113.26	&	383.08	&	189.04	&	109	\\
	&	137.06	&	494.06	&	246.90	&	137.48	&	495.74	&	247.77	&	138.16	&	498.64	&	249.27	&	139.34	&	501.67	&	250.83	&	137	\\
	&	217.02	&	1010.68	&	511.62	&	217.09	&	1010.97	&	511.78	&	217.22	&	1012.28	&	512.46	&	217.60	&	1014.69	&	513.71	&	217	\\
7.9	&	32.58	&	80.44	&	45.29	&	48.38	&	117.50	&	61.25	&	59.39	&	150.10	&	74.43	&	68.98	&	179.06	&	86.67	&	27	\\
	&	37.31	&	95.08	&	51.95	&	50.16	&	126.26	&	64.78	&	60.34	&	155.69	&	76.74	&	69.59	&	182.81	&	88.30	&	34	\\
	&	44.83	&	120.47	&	62.44	&	53.89	&	143.95	&	71.92	&	62.61	&	168.05	&	81.93	&	71.11	&	192.05	&	92.40	&	43	\\
	&	55.06	&	158.62	&	77.96	&	60.71	&	173.66	&	84.33	&	67.30	&	190.77	&	91.83	&	74.43	&	209.41	&	100.32	&	54	\\
	&	68.68	&	209.94	&	100.58	&	71.77	&	218.12	&	104.41	&	76.11	&	228.40	&	109.32	&	81.44	&	240.94	&	115.42	&	68	\\
	&	86.38	&	272.52	&	131.26	&	87.81	&	277.19	&	133.65	&	90.87	&	282.23	&	137.19	&	94.62	&	290.64	&	140.56	&	86	\\
	&	109.11	&	356.83	&	175.23	&	109.92	&	359.09	&	176.42	&	111.21	&	362.70	&	178.23	&	113.01	&	368.99	&	181.63	&	109	\\
	&	137.09	&	476.76	&	237.96	&	137.59	&	478.66	&	238.94	&	138.34	&	481.08	&	240.19	&	139.35	&	484.79	&	242.11	&	137	\\
	&	217.02	&	974.45	&	492.82	&	217.09	&	975.07	&	493.14	&	217.27	&	975.94	&	493.59	&	217.59	&	978.74	&	495.04	&	217	\\
10.0	&	32.54	&	77.98	&	44.08	&	48.32	&	113.63	&	59.68	&	59.28	&	145.16	&	72.41	&	68.84	&	173.51	&	84.27	&	27	\\
	&	37.28	&	91.67	&	50.46	&	50.05	&	121.79	&	62.98	&	60.22	&	150.46	&	74.58	&	69.44	&	176.97	&	85.76	&	34	\\
	&	44.82	&	116.07	&	60.67	&	53.85	&	138.64	&	69.76	&	62.50	&	162.17	&	79.44	&	70.96	&	185.53	&	89.50	&	43	\\
	&	55.05	&	152.70	&	75.50	&	60.60	&	167.19	&	81.56	&	67.20	&	183.96	&	88.81	&	74.34	&	202.37	&	97.08	&	54	\\
	&	68.66	&	202.10	&	96.96	&	71.72	&	210.16	&	100.68	&	76.06	&	220.01	&	105.32	&	81.38	&	231.67	&	110.90	&	68	\\
	&	86.37	&	262.93	&	126.39	&	88.03	&	267.95	&	128.93	&	90.55	&	272.98	&	131.49	&	94.17	&	279.81	&	134.99	&	86	\\
	&	109.12	&	343.60	&	168.27	&	110.16	&	346.52	&	169.80	&	111.18	&	351.48	&	172.41	&	113.23	&	355.11	&	174.32	&	109	\\
	&	137.10	&	458.98	&	228.73	&	137.62	&	460.71	&	229.63	&	138.16	&	463.29	&	230.97	&	139.31	&	466.40	&	232.58	&	137	\\
	&	217.02	&	938.07	&	474.00	&	217.09	&	938.65	&	474.30	&	217.22  &	939.88	&	474.93	&	217.59	&	942.07	&	476.06	&	217	\\
12.6	&	32.53	&	75.61	&	42.88	&	48.26	&	109.92	&	58.17	&	59.17	&	140.49	&	70.52	&	68.72	&	168.23	&	82.01	&	27	\\
	&	37.27	&	88.63	&	49.11	&	49.98	&	117.66	&	61.31	&	60.13	&	145.40	&	72.51	&	69.31	&	171.54	&	83.42	&	34	\\
	&	44.83	&	111.89	&	58.97	&	53.78	&	133.81	&	67.81	&	62.41	&	156.53	&	77.09	&	70.84	&	179.42	&	86.83	&	43	\\
	&	55.03	&	146.97	&	73.15	&	60.60	&	161.00	&	78.95	&	67.20	&	176.78	&	85.68	&	74.22	&	195.32	&	93.88	&	54	\\
	&	68.45	&	193.69	&	93.14	&	71.70	&	202.18	&	97.00	&	76.04	&	212.20	&	101.63	&	81.23	&	224.10	&	107.26	&	68	\\
	&	86.23	&	252.90	&	121.35	&	87.95	&	257.84	&	123.82	&	90.52	&	263.30	&	126.57	&	94.33	&	271.78	&	130.88	&	86	\\
	&	109.12	&	331.47	&	161.88	&	110.09	&	334.51	&	163.48	&	111.28	&	337.63	&	165.13	&	113.00	&	342.70	&	167.79	&	109	\\
	&	137.06	&	442.53	&	220.17	&	137.48	&	444.04	&	220.96	&	138.30	&	446.57	&	222.28	&	139.29	&	449.89	&	224.01	&	137	\\
	&	217.02	&	902.64	&	455.74	&	217.09	&	903.60	&	456.24	&	217.26	&	904.67	&	456.79	&	217.59	&	906.90	&	457.93  & 	217	\\

\enddata
\tablecomments{Surface gravities are in m s$^{-2}$.  A metallicity of [M/H]=+1.5 ($\sim$~30$\times$ solar) is assumed.  ``0.12N'' and ``1.0N'' mean 0.12 and 1.0 times the current Neptunian incident flux, while ``1.0U'' and ``1.8U'' mean 1.0 and 1.8 times the current Uranian incident flux.  The temperatures at 1 and 10 bars are provided.}
\label{untab}
\end{deluxetable}
\end{center}